\documentclass[twocolumn]{aastex63}

\newcommand\gaia{\textit{Gaia}}
\newcommand\kms{$\textrm{km/s}$}

\newcommand\teff{T$_{\rm{eff}}$}
\newcommand\vsini{$v$~sin~$i$}

\newcommand\earthmass{$M_{\oplus}$}
\newcommand\earthradius{$R_{\oplus}$}

\newcommand\specificflux{ergs $\rm{s^{-1} cm^{-2}} \AA^{-1}$}
\newcommand\flux{ergs $\rm{s^{-1} cm^{-2}}$}

\newcommand\halpha{H$\alpha$}

\shortauthors{Kanodia et al. 2021}
\shorttitle{NIR observations of a flaring vB 10}

\received{August 31, 2021}
\accepted{November 25, 2021}
\submitjournal{ApJ}

\begin{document}

\title{High resolution near-infrared spectroscopy of a flare around the ultracool dwarf - vB 10}

\author[0000-0001-8401-4300]{Shubham Kanodia}
\affiliation{Department of Astronomy \& Astrophysics, 525 Davey Laboratory, The Pennsylvania State University, University Park, PA, 16802, USA}
\affiliation{Center for Exoplanets and Habitable Worlds, 525 Davey Laboratory, The Pennsylvania State University, University Park, PA, 16802, USA}

\author[0000-0002-4289-7958]{Lawrence W. Ramsey}
\affiliation{Department of Astronomy \& Astrophysics, 525 Davey Laboratory, The Pennsylvania State University, University Park, PA, 16802, USA}
\affiliation{Center for Exoplanets and Habitable Worlds, 525 Davey Laboratory, The Pennsylvania State University, University Park, PA, 16802, USA}

\author[0000-0001-8222-9586]{Marissa Maney}
\affil{Department of Astronomy \& Astrophysics, 525 Davey Laboratory, The Pennsylvania State University, University Park, PA, 16802, USA}
\affil{Center for Exoplanets and Habitable Worlds, 525 Davey Laboratory, The Pennsylvania State University, University Park, PA, 16802, USA}

\author[0000-0001-9596-7983]{Suvrath Mahadevan}
\affil{Department of Astronomy \& Astrophysics, 525 Davey Laboratory, The Pennsylvania State University, University Park, PA, 16802, USA}
\affil{Center for Exoplanets and Habitable Worlds, 525 Davey Laboratory, The Pennsylvania State University, University Park, PA, 16802, USA}

\author[0000-0003-4835-0619]{Caleb I. Ca\~nas}
\affil{Department of Astronomy \& Astrophysics, 525 Davey Laboratory, The Pennsylvania State University, University Park, PA, 16802, USA}
\affil{Center for Exoplanets and Habitable Worlds, 525 Davey Laboratory, The Pennsylvania State University, University Park, PA, 16802, USA}
\affiliation{NASA Earth and Space Science Fellow}

\author[0000-0001-8720-5612]{Joe P. Ninan}
\affil{Department of Astronomy \& Astrophysics,  525 Davey Laboratory, The Pennsylvania State University,  University Park, PA, 16802, USA}
\affil{Center for Exoplanets and Habitable Worlds, 525 Davey Laboratory, The Pennsylvania State University, University Park, PA, 16802, USA}

\author[0000-0002-0048-2586]{Andrew Monson}
\affil{Department of Astronomy \& Astrophysics, 525 Davey Laboratory, The Pennsylvania State University, University Park, PA, 16802, USA}
\affil{Center for Exoplanets and Habitable Worlds, 525 Davey Laboratory, The Pennsylvania State University, University Park, PA, 16802, USA}

\author[0000-0001-7458-1176]{Adam F. Kowalski}
\affil{Department of Astrophysical and Planetary Sciences, University of Colorado Boulder, 2000 Colorado Ave,
Boulder, CO 80305, USA}
\affil{National Solar Observatory, University of Colorado Boulder, 3665 Discovery Drive, Boulder, CO 80303, USA}
\affil{Laboratory for Atmospheric and Space Physics, University of Colorado Boulder, 3665 Discovery Drive, Boulder, CO 80303, USA}

\author{Maximos C. Goumas}
\affil{Department of Aerospace, Physics, and Space Sciences, Florida Institute of Technology, Melbourne, FL 32901, USA}

\author[0000-0001-7409-5688]{Gudmundur Stefansson}
\affiliation{Henry Norris Russell Fellow}
\affiliation{Department of Astrophysical Sciences, Princeton University, 4 Ivy Lane, Princeton, NJ 08540, USA}

\author[0000-0003-4384-7220]{Chad F. Bender}
\affil{Steward Observatory, The University of Arizona, 933 N.\ Cherry Avenue, Tucson, AZ 85721, USA}

\author[0000-0001-9662-3496]{William D. Cochran}
\affiliation{Department of Astronomy, The University of Texas at Austin, TX 79734, USA}
\affiliation{McDonald Observatory, The University of Texas at Austin, TX 79734, USA}
\affiliation{Center for Planetary Systems Habitability, The University of Texas at Austin, USA}

\author[0000-0002-2144-0764]{Scott A. Diddams}
\affil{Time and Frequency Division, National Institute of Standards and Technology, 325 Broadway, Boulder, CO 80305, USA}
\affil{Department of Physics, University of Colorado, 2000 Colorado Avenue, Boulder, CO 80309, USA}

\author[0000-0002-0560-1433]{Connor Fredrick}
\affil{Time and Frequency Division, National Institute of Standards and Technology, 325 Broadway, Boulder, CO 80305, USA}
\affil{Department of Physics, University of Colorado, 2000 Colorado Avenue, Boulder, CO 80309, USA}

\author[0000-0003-1312-9391]{Samuel Halverson}
\affil{Jet Propulsion Laboratory, 4800 Oak Grove Drive, Pasadena, CA 91109, USA}

\author[0000-0002-1664-3102]{Fred Hearty}
\affil{Department of Astronomy \& Astrophysics, 525 Davey Laboratory, The Pennsylvania State University, University Park, PA, 16802, USA}
\affil{Center for Exoplanets and Habitable Worlds, 525 Davey Laboratory, The Pennsylvania State University, University Park, PA, 16802, USA}

\author[0000-0001-9165-8905]{Steven Janowiecki}
\affiliation{McDonald Observatory, The University of Texas at Austin, TX 79734, USA}

\author[0000-0001-5000-1018]{Andrew J. Metcalf}
\affiliation{Space Vehicles Directorate, Air Force Research Laboratory, 3550 Aberdeen Ave. SE, Kirtland AFB, NM 87117, USA}

\author{Stephen C. Odewahn}
\affiliation{Department of Astronomy, The University of Texas at Austin, TX 79734, USA}
\affiliation{McDonald Observatory, The University of Texas at Austin, TX 79734, USA}

\author[0000-0003-0149-9678]{Paul Robertson}
\affiliation{Department of Physics \& Astronomy, University of California Irvine, Irvine, CA 92697, USA}

\author[0000-0001-8127-5775]{Arpita Roy}
\affiliation{Space Telescope Science Institute, 3700 San Martin Dr, Baltimore, MD 21218, USA}
\affiliation{Department of Physics and Astronomy, Johns Hopkins University, 3400 N Charles Street, Baltimore, MD 21218, USA}

\author[0000-0002-4046-987X]{Christian Schwab}
\affil{Department of Physics and Astronomy, Macquarie University, Balaclava Road, North Ryde, NSW 2109, Australia}

\author[0000-0002-4788-8858]{Ryan C. Terrien}
\affil{Department of Physics and Astronomy, Carleton College, One North College Street, Northfield, MN 55057, USA}

\correspondingauthor{Shubham Kanodia}
\email{shbhuk@gmail.com}


\begin{abstract}
We present high-resolution observations of a flaring event in the M8 dwarf vB 10 using the near-infrared Habitable zone Planet Finder (HPF) spectrograph on the Hobby Eberly Telescope (HET).  The high stability of HPF enables us to accurately subtract a VB 10 quiescent spectrum from the flare spectrum to isolate the flare contributions, and study the changes in the relative energy of the Ca II infrared triplet (IRT), several Paschen lines, the He 10830 \AA~ triplet lines, and select iron and magnesium lines in HPF’s bandpass. Our analysis reveals the presence of a red asymmetry in the He 10830 \AA~  triplet; which is similar to signatures of coronal rain in the Sun. Photometry of the flare derived from an acquisition camera before spectroscopic observations, and the ability to extract spectra from up-the-ramp observations with the HPF infrared detector, enables us to perform time-series analysis of part of the flare, and provide coarse constraints on the energy and frequency of such flares. We compare this flare with historical observations of flares around vB 10 and other ultracool M dwarfs, and attempt to place limits on flare-induced atmospheric mass loss for hypothetical planets around vB 10.

\end{abstract}


\section{Introduction} \label{sec:intro}
Stellar flares are a common phenomenon around M dwarfs. These transient events take place when magnetic field lines reconnect in the upper atmosphere of the star. This converts a large amount of magnetic energy to kinetic energy of electrons, and ions, which is then redistributed across the electromagnetic spectrum. These non-thermal accelerated electrons collide with the cold, thick lower chromosphere and upper photosphere, and emit a white light continuum, which is approximately a blackbody of temperature 10,000 K \citep{mochnacki_multichannel_1980, kahler_coordinated_1982, kowalski_new_2015}. This blackbody emission dominates in the blue/visible over the M dwarf continuum, and is typically accompanied with high energy X-ray and UV radiation.  

Flares around M dwarfs have been studied photometrically since the 1940s \citep{van_maanen_photographic_1940, joy_observations_1949}, with the first spectroscopic observations following soon after \citep{herbig_observations_1956}. \cite{kunkel_spectra_1970} presented a two-component spectral model consisting of hydrogen recombination similar to Solar flares, as well as a second component which follows shortly after: an impulsive component that heats the photosphere. However, this model does not include the spectral features observed during flares such as line emission. While the continuum emission is typically that of a blackbody of $\sim 10,000$ K, the emission lines can include H, He, and Ca ionic species at chromospheric temperatures\footnote{For an introduction to M dwarf flares, refer to \cite{reid_new_2000}.}.

Studies of M dwarf flares using high resolution spectroscopy are sparse due to the challenges associated with their faintness, which is an even bigger problem for ultracool dwarfs. High resolving power ($R \sim 110,000$) spectra were obtained for the M7 dwarf vB 8 by \cite{martin_utrecht_1999} covering a bandpass spanning about 5400 \AA~-- 10600 \AA. These observations showed the rise and fall of the He I D3 line as well as the Na I doublet during the flare of this cool dwarf.  \cite{crespo-chacon_analysis_2006} obtained medium resolution optical spectra (3500 \AA~-- 7200 \AA) for the active M3 star AD Leo to study the chromospheric lines during multiple flares.  High resolution ($R \sim 60,000$) optical spectra (3600 \AA~-- 10800 \AA) were obtained for the M4 dwarf GJ 699  \citep[Barnard's star;][]{paulson_optical_2006}, which showed Stark broadening for the Balmer lines. \cite{schmidt_activity_2007} observed a flare for the M7 dwarf 2MASS J1028404--143843 in the red-optical (6000 \AA~-- 10000 \AA) covering \halpha{}, and some of the Paschen lines \deleted{, which are discussed further in subsequent sections}. \cite{fuhrmeister_multiwavelength_2008} studied a flare around the M5.5 dwarf CN Leo using observations spanning 3000\AA~-- 10500 \AA, at $R \sim 40,000$. These observations identified not just a range of chromospheric lines in emission, but also continuum enhancement and line asymmetry. \cite{fuhrmeister_simultaneous_2007} obtained simultaneous observations of flaring activity around the M5.5 dwarf CN Leo spanning 3000\AA~-- 10500 \AA, at $R \sim 40,000$, as well as X-ray data from \textit{XMM Newton}.
\cite{fuhrmeister_multi-wavelength_2011} observed a flare in the optical at ($R \sim 45,000$) around Proxima Centauri, the closest star to the Sun, which was contemporaneous with observations in the X-ray. Using these measurements, they also discuss theoretical models to constraint the chromospheric properties of the star during the flare. \cite{honda_time_2018} reported observations of a stellar flare around  EV Lac (M4) spanning 6350\AA~-- 6800 \AA~ at $R \sim 10,000$, covering \halpha{} and the He I lines. They studied the line asymmetries associated with the wings of  \halpha{} associated with different stages of the flare. \cite{kowalski_near-ultraviolet_2019} reported NUV (2444\AA~ -- 2841\AA) spectroscopic observations of the M4 dwarf GJ 1243, which showed an increase in the broadband continuum during flare events. Recently, \cite{muheki_high-resolution_2020} studied AD Leo at $R \sim 35,000$ from 4536\AA~-- 7592 \AA~ to gain insight into flares and coronal mass ejections (CMEs).

\cite{guedel_x-raymicrowave_1993} suggest that the plasma heating, and X-ray luminosities from particle accelerations during the quiescent phase should dramatically change around the M7 spectral subtype. Yet, the spectral properties of flares around ultracool dwarfs are similar to those around earlier M dwarfs. \cite{lacy_uv_1976} (for an updated plot, see \cite{osten_drafts_2012}), showed that the average flare energy release rate and average flare energy are correlated with the quiescent bolometric luminosity of the star, yet we see flares for ultracool dwarfs that release energy comparable to their bolometric luminosity. The growing suite of instruments \citep[HPF, CARMENES, SPIROU, MAROON-X, IRD, NIRPS; ][]{mahadevan_habitable-zone_2014, quirrenbach_carmenes_2014, seifahrt_development_2016, donati_spirou_2020, kotani_infrared_2018, wildi_nirps_2017} searching for planets around M dwarfs using high resolution doppler spectrographs in the optical and NIR, offers an increasing volume of spectra that is useful for serendipitous stellar flare observations \citep[e.g.,][]{fuhrmeister_carmenes_2020} to answer these open questions about flares around ultracool dwarfs.

Not only are M dwarfs the most common spectral type of stars in the Galaxy \citep{henry_solar_2006}, but also they typically host more than one rocky exoplanet \citep{dressing_occurrence_2015, mulders_stellar-mass-dependent_2015, hardegree-ullman_kepler_2019}. They present favourable targets for planet detection due to their smaller radii and masses compared to Solar type stars \citep{scalo_m_2007, irwin_mearth-north_2015}, and are also favourable for planetary atmospheric characterization \citep{batalha_precision_2019}. Understanding the occurrence and energetics of flares around M dwarfs is crucial not just from a stellar astrophysics perspective, but also to understand the stellar environment influencing the evolution of exoplanets and their atmospheres \citep{luger_habitable_2015, cuntz_about_2016, owen_habitability_2016, mcdonald_sub-neptune_2019}.

In this work, we present observations and analysis of high resolution NIR spectra of vB 10, which were obtained with HPF on the 10 m HET, where \replaced{two flares were}{a stellar flare was} serendipitously detected. In Section \ref{sec:vB 10}, we discuss the target and previous studies associated with its flaring activity. In Section \ref{sec:hpf}, we discuss the observations, \deleted{corrections} and processing performed on the spectra, while in Section \ref{sec:het_acam}, we discuss photometric observations that are almost contemporaneous with the spectra, and help us estimate the continuum enhancement during this flare. In Section \ref{sec:lines}, we detail the analysis performed on the spectral lines seen in the vB 10 flares, and place it in context of other flares seen around similar stars. In Section \ref{sec:discussion}, we estimate the bolometric luminosity of the flare, as well as its frequency. We also discuss implications of such flares on planetary atmosphere escape, and conclude in Section \ref{sec:conclusion}.

\section{vB 10}\label{sec:vB 10}

vB 10 (GJ 752 B, 2MASS J19165762+0509021), discovered by van Biesbroeck in 1944 \citep{van_biesbroeck_star_1944}, is a relatively slowly rotating \citep[\vsini{} $\sim  2.7 $ \kms{};][]{reiners_carmenes_2018}, fully convective M8 star \citep{liebert_spectrophotometry_1978, kirkpatrick_solar_1995} with a log (L$_{\rm{bol}}$/L$_{\rm{\odot}}$) of -3.36 \citep{tinney_faintest_1993}. 

Flares on very cool M dwarfs have been known since \cite{herbig_observations_1956} first discovered dramatic brightening in Balmer lines $\&$ Ca II H$\&$K in vB 10. It also has an extensive history of multi-wavelength flare observations:  \cite{linsky_stellar_1995} observed a flare in UV chromospheric and transition region lines with the Goddard High Resolution Spectrograph (GHRS) on the \textit{Hubble Space Telescope} (\textit{HST}).  \cite{fleming_x-ray_2000} first reported an X ray flare from vB 10, whereas \cite{berger_simultaneous_2008-1} conducted simultaneous X-ray, UV, and optical observations and observed several flares on vB 10.

\begin{deluxetable*}{ccccc}
\tablecaption{Summary of the HPF observations analyzed in this manuscript \label{tab:HPFobservations}}
\tablehead{\colhead{Object} & \colhead{JDUTC$^a$} & \colhead{S/N$^b$} & \colhead{Exp.Time} & \colhead{Comment}}
\startdata
vB 10    & 2019 Aug 20 05:40    & 105 & 945s      & Flare - T1      \\
vB 10    & 2019 Aug 20 05:57   & 119 & 945s      & Flare - T2      \\
vB 10    & 2018 Sept 24 03:20   & 132 & 945s      & Template          \\
vB 10    & 2018 Sept 24 03:36     & 144 & 945s      & Template          \\
HR 3437  & 2018 Nov 22 11:36    & 398 & 330s      & Inst. Response A0 \\
16 Cyg B & 2020 Mar 24 11:41   & 271 & 202s x5   & Inst. Response G3
\enddata
\tablenotetext{a}{Start of HPF exposure}
\tablenotetext{b}{Per 1D extracted pixel, calculated as the median S/N for HPF order index 18 ($\sim$ 10630 \AA~ -- 10770 \AA)}
\end{deluxetable*}

\section{HPF spectroscopy}\label{sec:hpf}
HPF \citep{mahadevan_habitable-zone_2012, mahadevan_habitable-zone_2014} is a NIR cross-dispersed echelle spectrograph ($R \sim 55,000)$ spanning 8100 - 12800 \AA. It is located at HET, McDonald Observatory, Texas, USA \citep{ramsey_early_1998}, and is an environmentally stabilized \citep{stefansson_versatile_2016}, fiber-fed spectrograph \citep{kanodia_overview_2018}. vB 10 was observed as part of the long term HPF  monitoring and guaranteed time of observation (GTO) program which began in  April 2018. Each visit included two individual exposures with a total exposure time of 945 s, in sampling up-the-ramp (SUTR) mode where each sub-frame has an integration time of 10.6 s, and all these non-destructive-readout (NDR) frames are combined to calculate a single 2D flux image on the detector. This allows us to reconstruct the change in the measured spectra as a function of time.

As part of these observations\footnote{Observations from April 2018 to October 2020 were analyzed for this work.}, spectra were obtained for vB 10 during a flare on 2019 August 20 \added{(JDUTC $\sim$ 2458715.75)} for 2x 945 s exposures starting 05:40 UTC \deleted{(hereafter referred to as Flare I)}. These observations are summarized in \autoref{tab:HPFobservations}.

We use the algorithms from \texttt{HxRGproc} \citep{ninan_habitable-zone_2018} for bias noise removal, nonlinearity correction, cosmic-ray correction, and slope/flux and variance image calculation  of the raw HPF data. This variance estimate is used to calculate the S/N of each HPF exposure (\autoref{tab:HPFobservations}). In addition, we flat correct the data, and extract it following the procedures in \cite{ninan_habitable-zone_2018},  and \cite{metcalf_stellar_2019}. After extracting the 2D spectrum to a 1D flux vs wavelength grid, we correct for telluric absorption as well as emission lines, and then shift the spectrum to the stellar rest frame. We then flux calibrate the spectrum, and correct for the chromatic instrument response. Finally, we subtract a quiescent phase spectrum from the flare spectrum to quantify the emission during the flare. The procedure for this is detailed in Section \autoref{appsec:spectralreduction}. All HPF spectra shown and discussed in this manuscript are in vacuum wavelengths.

\section{HET Acquisition camera (ACAM) photometry}\label{sec:het_acam}

\begin{figure*}[!t] 
\centering
\includegraphics[width=0.95\textwidth]{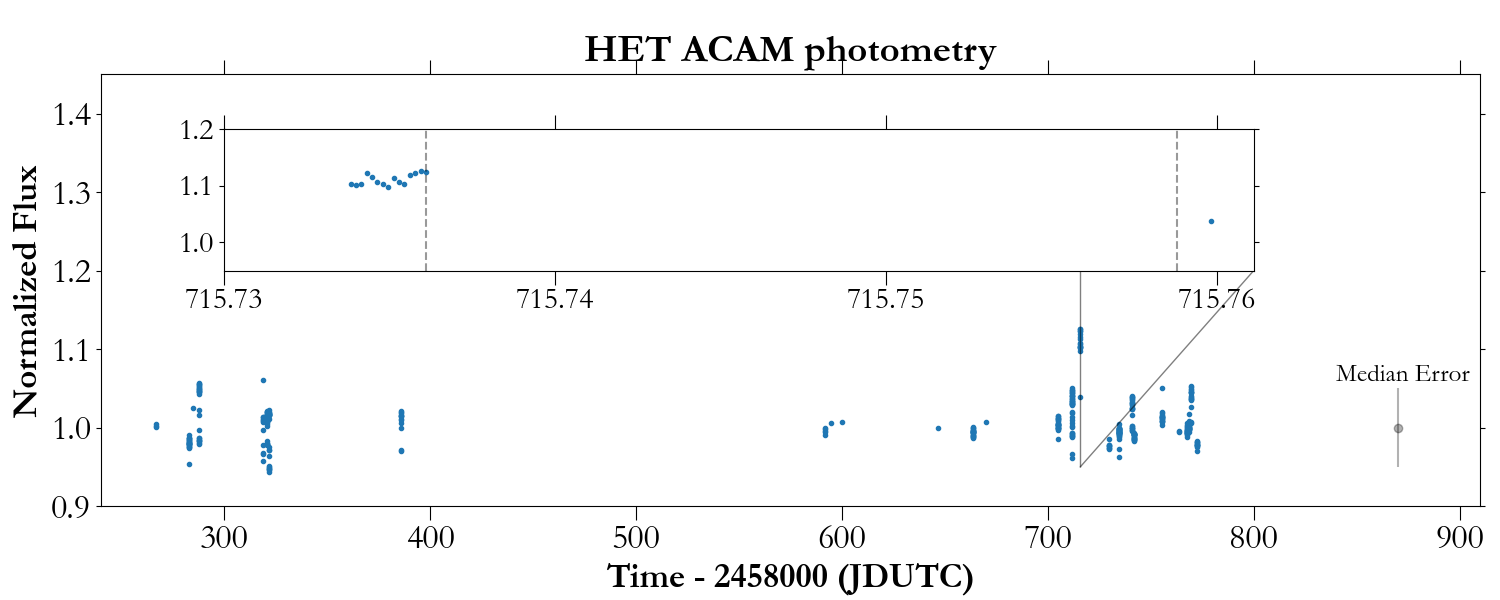}
\caption{Relative photometry of vB 10 as seen by HET ACAM, with a typical error of $\sim 5\%$ on each point. The inset shows the acquisition photometry preceding T1, with the span of the HPF observations marked with the dashed lines. We see a 12\% enhancement in the photometry preceding the HPF observing window (marked with the dashed lines), while our median error of $\sim 5\%$ is also plotted for representative purposes. Since the photometry is consistently elevated across 15 ACAM exposures to $\sim 12\%$ before T1, we do not attribute this enhancement to a statistical anomaly, and discuss the implications in Section \ref{sec:flareenergy}.} \label{fig:ACAM}
\end{figure*}

As part of the target acquisition at HET, vB 10 was observed on the HET Acquisition camera (ACAM), which has a field of view of 3.5$\arcmin$ $\times$ 3.5$\arcmin$. Typical HPF acquisition includes a few ACAM images before the HPF exposures to locate the target, and then move it to the HPF telescope fiber \citep{kanodia_overview_2018}, and also one ACAM image after the HPF exposure. The images were taken using the SDSS $i^\prime$ filter\footnote{Centered at 7718 \AA~ with a FWHM of 1564 \AA.} with exposure times varying between 3 -- 10 s, and $1 \times 1$ on-chip binning.  We use these ACAM images to perform differential aperture photometry on vB 10 and search for any enhancement in the broadband continuum during the flare. The procedure followed to reduce the photometry is detailed in \added{Appendix} Section \ref{appsec:acam}.

We show the photometry from the HET ACAM in \autoref{fig:ACAM}. An ACAM sequence of exposures spanning $\sim 3$ minutes was taken preceding \replaced{Flare I}{T1}, which ended under 1 minute before the start of the HPF exposure (Flare - T1; \autoref{tab:HPFobservations}). Additionally, one ACAM exposure was taken about 1 minute after the end of the HPF exposure (Flare - T2). We note that the fluxes during this sequence of exposures before T1 are about $\sim 12\%$ higher than the median normalized baseline over the entire observing sequence. The implications of this continuum enhancement are discussed in Section \ref{sec:flareenergy}. 


\section{Results for Activity Sensitive Features}\label{sec:lines}

\begin{deluxetable*}{lccccccc}
\tablecaption{List of the lines in emission that are analyzed \deleted{for Flare I}, with the fluxes measured as described in Section \ref{sec:lines}, with a typical uncertainty on the fluxes of $\sim 15 \%$. All wavelengths listed below are in {the stellar rest frame}{in vacuum, and corrected to the stellar rest frame}  \citep{kramida_critical_2010}. \deleted{The fluxes are in ergs s$^{-1}$ cm$^{-2}$}. A machine readable version of this table is included with this manuscript. \label{tab:line_flare1}}
\tablehead{\colhead{Atomic} & \colhead{HPF Order} & \multicolumn{2}{c}{Wavelength} & \multicolumn{2}{c}{Line Integral Flux}  &
\multicolumn{2}{c}{Luminosity}  \\
\colhead{} & \colhead{} & \multicolumn{2}{c}{(Angstrom)} & \multicolumn{2}{c}{($\times 10^{-14}$ ergs s$^{-1}$ cm$^{-2}$)} & \multicolumn{2}{c}{($\times 10^{25}$ ergs/s)} \\
\colhead{Line} & \colhead{Index$^a$} & \colhead{Vacuum} & \colhead{Air} & \colhead{T1} & \colhead{T2} &  \colhead{T1} & \colhead{T2}}
\startdata
Ca II & 3        & 8500.4  & 8498    & 3.01        & 2.08        & 12.04          & 8.32           \\
Ca II & 4        & 8544.4  & 8542.1  & 3.02        & 2.27        & 12.08          & 9.08          \\
Ca II & 5        & 8664.5  & 8662.1  & 2.31        & 1.87        & 9.24           & 7.48           \\
Fe I  & 5        & 8691    & 8688.6  & 0.208       & 0.152       & 0.83           & 0.61        \\
Pa 12$^b$ & 5        & 8752.9  & 8750.3  & 0.296       & 0.224       & 1.18           & 0.90          \\
Mg I  & 6        & 8809.2  & 8806.8  & 0.215       & 0.248       & 0.86           & 0.99     \\
Fe I  & 6        & 8826.6  & 8824.2  & 0.214       & 0.213       & 0.86           & 0.85              \\
Pa 11$^b$  & 6        & 8865.2  & 8862.9  & 0.524       & 0.402       & 2.10           & 1.61          \\
Pa 7 ($\delta$)$^b$  & 14       & 10052.1 & 10049.4 & 1.48        & 0.999      & 5.92           & 4.00         \\
He I$^{b}$  & 19       & 10832.1 & 10829.1 & 1.28        & 1.1                             & 5.12          & 4.4             \\
He I$^{b}$  & 19       & 10833.2 & 10830.2 & 1.54        & 1.42                             & 6.16          & 5.68            \\
He I$^{b}$  & 19       & 10833.3 & 10830.3 & 1.54        & 1.42                             & 6.16          & 5.68              \\
Pa 6 ($\gamma$)$^b$  & 19       & 10941.1 & 10938.1 & 1.75        & 1.08                         & 7.00           & 4.32            
\enddata
\tablenotetext{a}{Zero indexed HPF orders}
\tablenotetext{b}{Lorentzian Fit}
\end{deluxetable*}

In this section we discuss the fluxes measured from the subtracted spectrum of the activity sensitive features during the \replaced{two flares (Flare I and II)}{stellar flare} observed in vB 10 with HPF. Optimized to obtain precise radial velocities on mid-to-late M dwarfs, the HPF bandpass covers the Ca II infrared triplet (IRT), as well as the He I 10830 \AA~transitions. Additionally, it spans the wavelength range of Pa $\gamma$ (Pa 6) through the Paschen jump. That said, most of the Paschen lines are either contaminated by water vapour, or in the case of the higher Paschen transitions, not seen. We also see several atomic lines in emission during the flare. 

All the spectra discussed hereafter are telluric corrected (Section \ref{sec:telluric}), shifted to the stellar rest frame (Section \ref{sec:velocity}),  response corrected (Section \ref{sec:relative}), flux calibrated (Section \ref{sec:absolute}), and quiescent phase subtracted (Section \ref{sec:spectralsubtraction}). This enables us to measure the energy emitted in these spectral features during the flare. In \autoref{tab:line_flare1}, we list the lines we analyzed, and also the measured fluxes for these lines during the two HPF observations \deleted{for Flare I}. The fluxes are measured by performing a Gaussian fit to calcium, iron and magnesium lines, and a Lorentzian fit to the Paschen and Helium lines. We propagate the error from the raw spectra, and calculate a final statistical error for the line fluxes to be $\sim 1\%$, which is consistent with the S/N of the observed spectra $\sim 100$ (per 1D extracted pixel, calculated as the median S/N for HPF order index 18 ($\sim$ 10630 \AA~ -- 10770 \AA)). (\autoref{tab:HPFobservations}). However to be conservative we also ascribe a systematic error of $15 \%$ to the flux estimates based on the response correction (Section \ref{sec:relative}). 

\replaced{For Flare I w}{W}e also compare the central wavelength from the line fit to the rest wavelength \citep{kramida_critical_2010}, and find the fits consistent with zero velocity offsets, with an absolute upper limit equivalent to the instrument resolution at $\sim 5$ \kms{}. This indicates the absence of bulk flowing material during the flare. This is consistent with the interpretation that these observations are during the decay phase of \replaced{Flare I}{the flare} (Section \ref{sec:ca_irt}). \replaced{Unlike \cite{fuhrmeister_multiwavelength_2008, fuhrmeister_multi-wavelength_2011}, and \cite{honda_time_2018}, we do not find any asymmetries in the Pa or Ca lines. We note the presence of a red asymmetry seen in the He triplet in Section \ref{sec:hetriplet}.}{Unlike \cite{fuhrmeister_multiwavelength_2008, fuhrmeister_multi-wavelength_2011}, and \cite{honda_time_2018}, we do not find any asymmetries in the Ca lines, while noting a weak red asymmetry excess for the Pa 6 ($\gamma$) line). Additionally, we discuss a red asymmetry seen in the He triplet in Section \ref{sec:hetriplet}.} In addition, we also reconstruct the temporal evolution of the fluxes across each 945 s exposure by using the SUTR sub frames across 3x intervals of 315 s each. 
 
There are very few reported observations to date of flares on ultracool M dwarfs at high spectral resolution, and even fewer that include the NIR HPF bandpass. \cite{schmidt_probing_2012} report on some of the first NIR observations using TripleSpec at the ARC 3.5 m telescope, but have very limited overlap with HPF, and report no IR data on vB 8, their latest spectral type target. The closest comparison to our vB 10 spectrum comes from flare observations reported by \cite{liebert_2mass_1999} for 2MASSW J0149090+295613 (an M9.5 dwarf, hereafter referred to as 2M0149), as well as from \cite{schmidt_activity_2007} for 2MASS J1028404--143843 (an M7 dwarf, hereafter referred to as 2M1028). We contrast our HPF vB 10 observations with these two datasets that span a bandpass from blue-wards of \halpha{} to about 1 $\mu$m, albeit at a much lower spectral resolution. In addition, they also bracket vB 10 (M8) in spectral type.

\subsection{Ca Infrared Triplet}\label{sec:ca_irt}

\begin{figure*}[!t] 
\centering
\includegraphics[width=\textwidth]{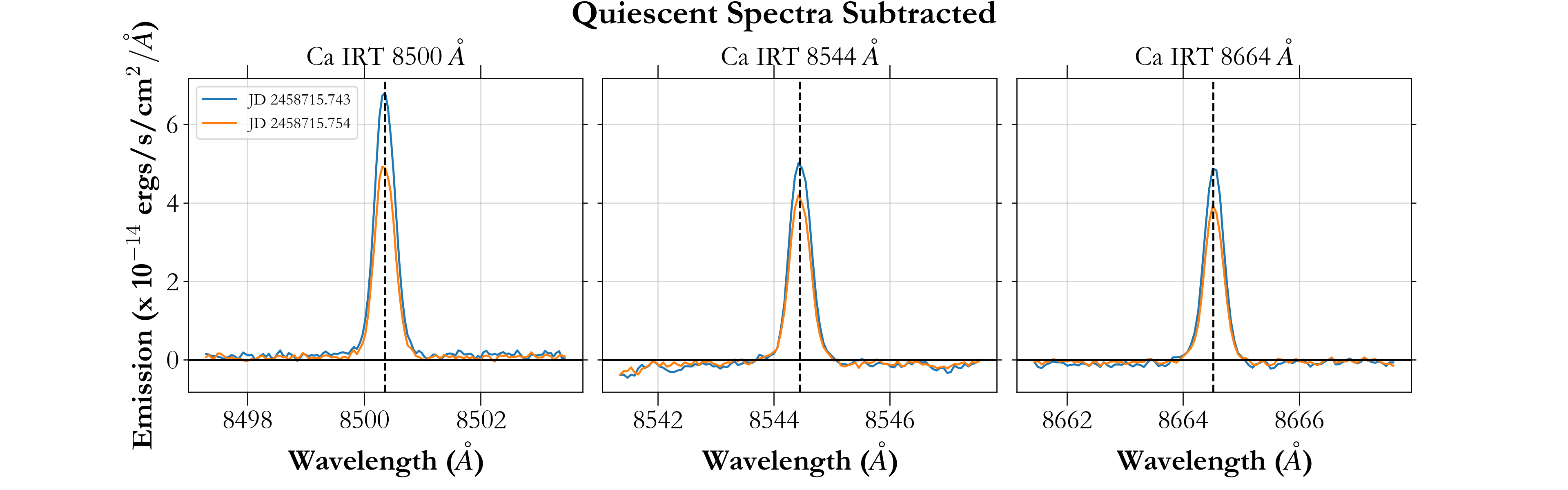}
\caption{Quiescent subtracted spectrum from HPF showing the emission in the Ca IRT seen in vB 10 \deleted{during flare I}. The dashed lines mark the rest position of each line. The two different colours indicate the successive HPF observations.} \label{fig:CaIRT_flare1}
\end{figure*}

\begin{deluxetable}{cccc}
\tablecaption{\gaia{} EDR3 parallax measurements \citep{collaboration_gaia_2021}\label{tab:parallaxes}}
\tablehead{\colhead{Name} & \colhead{\gaia{} EDR3 ID} & \colhead{Parallax} & \colhead{Distance}\\
\colhead{} & \colhead{} & \colhead{mas} & \colhead{parsec}}
\startdata
vB 10 & 4293315765165489536 & 168.9 & 5.92\\
2M0149 & 302525062400087936 & 42.4 & 23.6\\
2M1028 & 3752240939122060160 & 28.9 & 34.6\\
\enddata
\end{deluxetable}

The Ca II IRT lines are a relatively well studied activity indicator in M dwarfs as well as other stars with solar like magnetic activity. \cite{martin_ca_2017} measure the flux excess for a number of G and K stars after subtracting the spectrum of an inactive template star. They demonstrate that the IRT is well correlated with the traditional activity indicators based on Ca II H$ \&$ K lines, such as R'$_{\rm{HK}}$ and S$_{\rm{MWO}}$. Based on optical spectra for the M5.5 dwarf Proxima Centauri, \cite{robertson_proxima_2016} show a correlation between the IRT and \halpha{}. \cite{schofer_carmenes_2019} also use the spectral subtraction methods with the extensive CARMENES \citep{quirrenbach_carmenes_2014} M dwarf dataset to study the correlation between the IRT and other lines. The IRT is the strongest activity indicator in the HPF bandpass.  The quiescent subtracted spectrum for the IRT is shown in \autoref{fig:CaIRT_flare1}.

We also compare the flux emitted in the IRT in vB 10 with 2M0149 and 2M1028 for the strongest line 8544.4 \AA~. Normalized for distance (\autoref{tab:parallaxes}), the flux in 2M1028 appears to be $54 \times$ stronger, and the emission seen in 2M0149 is about $2 \times$ stronger. The ratio of the IRT lines, and particularly of 8544.4 \AA, the strongest, to 8500.4 \AA, the weakest, is a useful indicator of \replaced{the type of}{stellar} activity. This ratio is typically 9.1 in solar prominences \citep{landman_measurements_1977}, and 1.7 in a solar flare observed with the Lick Hamilton Echelle \citep{johns-krull_hamilton_1997}. In active binaries, the ratio of residual emission in the 8544 line to the 8500 line is typically 1.2 to 1.6 \citep{hall_eclipse_1992, montes_multiwavelength_2000}.

The \replaced{Flare I}{flare} observations yield a line ratio for 8544/8500 $\sim 1$ for the first observation which increases to 1.1 for the second. This indicates the flare was optically thick during this phase, while softening. This is further corroborated by the decreasing fluxes as discussed in Section \ref{sec:linedecay}. The ratio of the other IRT lines 8544/8665 is $\sim1.3$ for the first and 1.2 for the second observation.  We note that it is un-physical for the ratio of 8544/8500 to be less than the 8544/8665 ratio\footnote{According to \cite{landman_measurements_1977}, the ratio of intensities for the IRT when optically thin should be 1/9 : 1 : 1/2, respectively. Therefore, the 8662 \AA~line should be 4-5 x stronger than the 8500 \AA~line when optically thin; conversely under optically thick conditions they should be of similar intensities.}, however these ratios are consistent given the flux uncertainty of $\sim 15\%$ detailed in Section \ref{sec:relative}, especially in Order 3 with the instrument response asymmetry.

For 2M1028, the ratios of 8544/8500, and 8544/8665 are $\sim1.36$ and $\sim1.34$ respectively. Similarly, for 2M0149, the ratios are $\sim1.48$ and $\sim1.32$ respectively. It is clear from the IRT fluxes, that the material creating the excess emission in vB 10 is optically thick, and the decaying nature (both in strength and optical thickness\footnote{As the material that is creating the excess emission reduces in density with time, the line ratios increase, thereby tending towards being optically thin; simultaneously the line fluxes are also diminishing as shown in \autoref{fig:Lineevolution}.}) supports that it is a flare event.

\begin{figure*}[!t] 
\centering
\includegraphics[width=\textwidth]{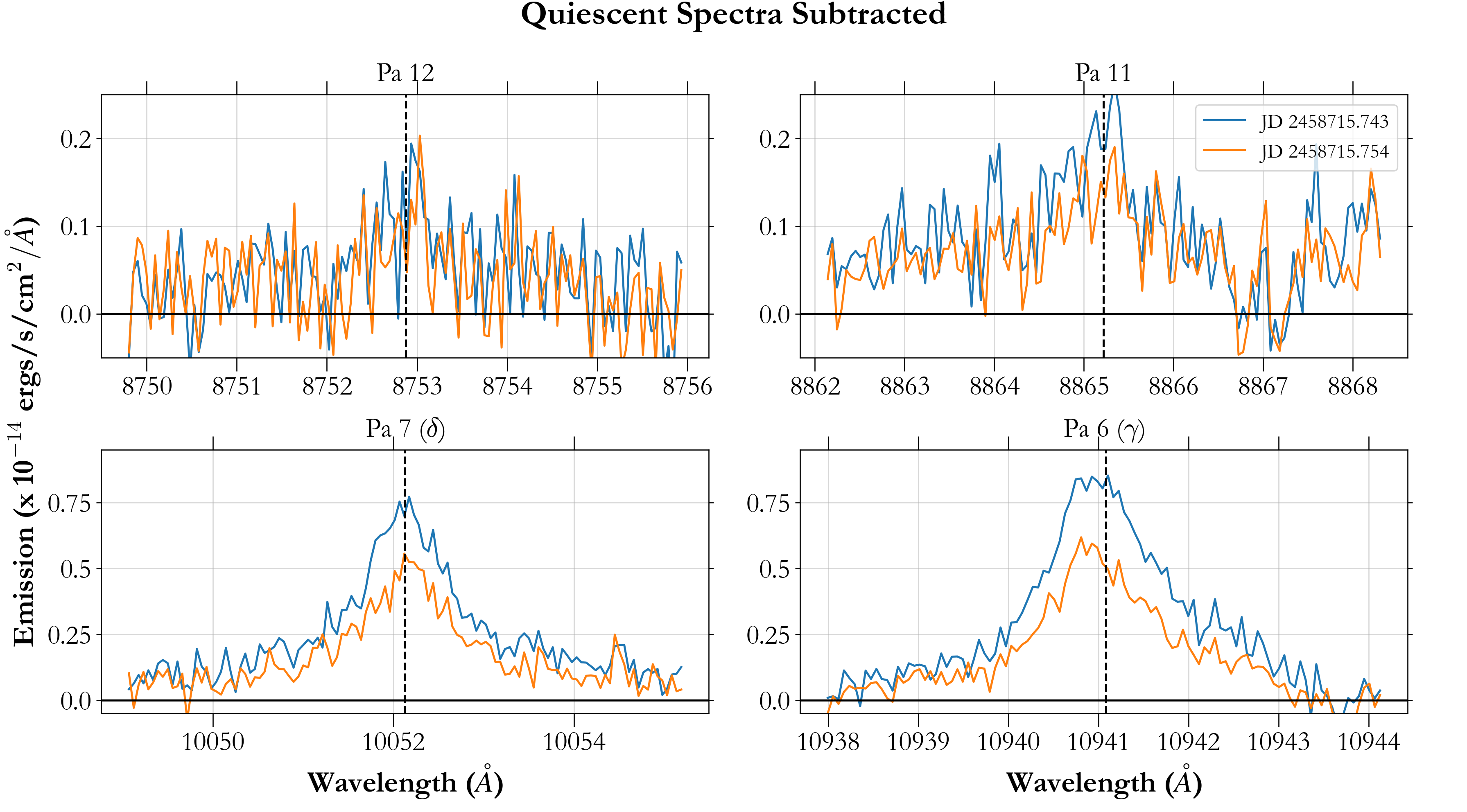}
\caption{Quiescent subtracted spectrum from HPF showing the Pa lines seen in emission in vB 10 \deleted{ during flare I}. The dashed lines mark the rest position of each line. Blue and orange depict the two observations --- T1 and T2 respectively.} \label{fig:Pa_flare1}
\end{figure*}

\subsection{Paschen lines}\label{sec:paschen}

There are several Paschen lines which fall within the HPF bandpass, of which Pa 12, Pa 11, Pa 7 ($\delta$), and Pa 6 ($\gamma$)\footnote{Pa 5 ($\beta$) falls just redwards of the HPF bandpass.} are measured to be in emission during \replaced{Flare I}{the flare} without significant telluric contamination (\autoref{tab:line_flare1}). The aforementioned studies for 2M1028 and 2M0149, as for the IRT, remain the best comparisons. Observations of Paschen emission in a Solar White-Light flare from April 1981 are discussed by \cite{neidig_hydrogen_1984}, and are in the context of detecting the Paschen continuum. \autoref{fig:Pa_flare1} shows the HPF vB 10 quiescent subtracted spectrum for these Pa lines during \replaced{Flare I}{the flare}. We note the broad wings present in the Paschen lines, which resemble the broad Balmer lines observed in a flare around GJ 699 by \cite{paulson_optical_2006}, which is indicative of linear Stark broadening \citep{paulson_optical_2006}.

\begin{figure}[!t] 
\centering
\includegraphics[width=\columnwidth]{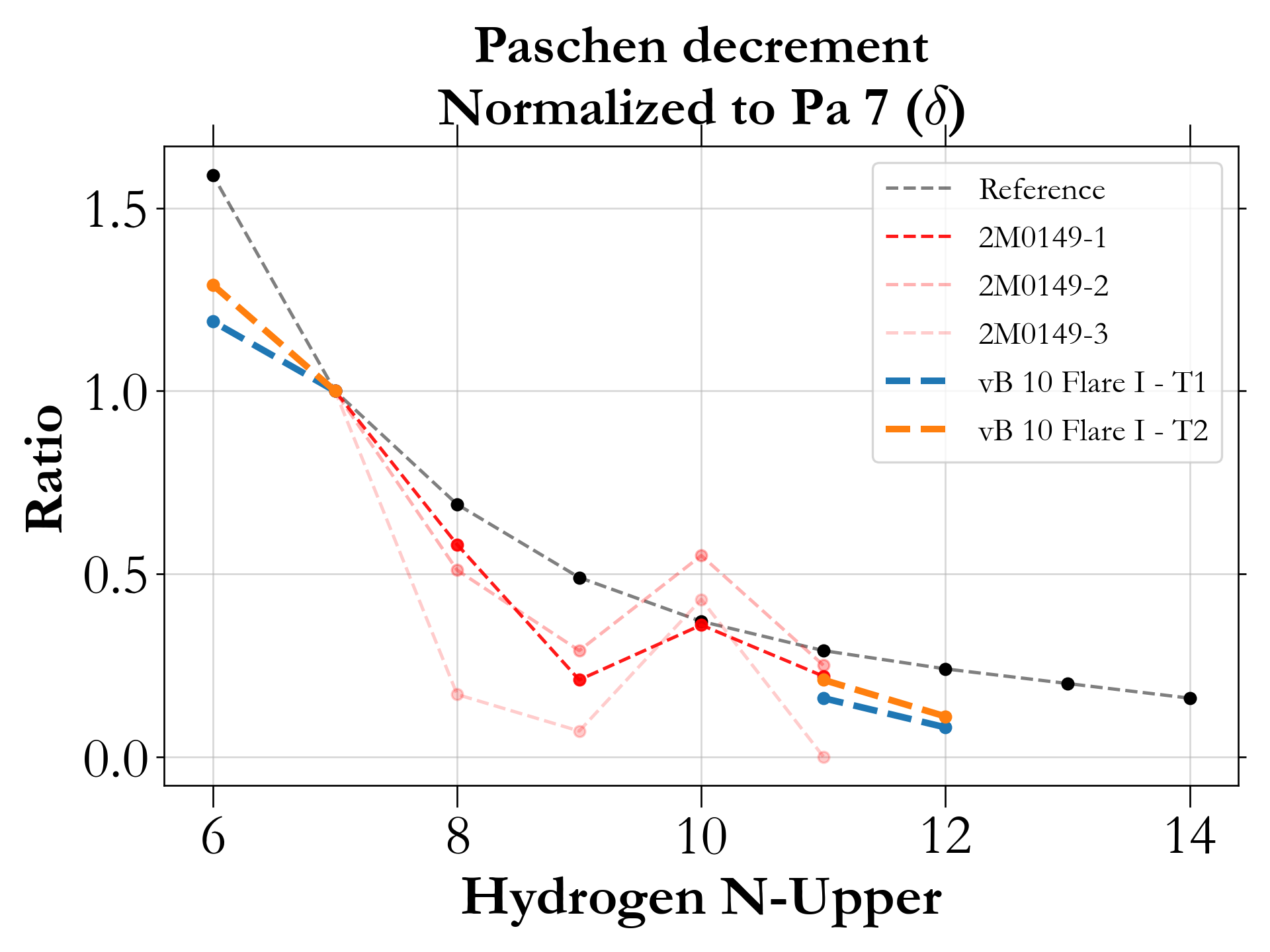}
\caption{The relative decrement of the vB 10 spectrum (blue and orange), compared to the 2M0149 observations (red). The reference line (black) is the Pa line decrement \citep{osterbrock_astrophysics_1989} normalized to Pa 7. } \label{fig:PaschenDec}
\end{figure}

The temporal decay for the Pa 6 line is discussed in Section \ref{sec:linedecay}. We also look at the relative decrement in the Paschen lines, plotting the normalized fluxes of the lines vs. the quantum number of the upper level of the Paschen transition (\autoref{fig:PaschenDec}). We choose to normalize with respect to Pa 7 ($\delta$) in this case because it is the lowest level that is common to vB 10 and 2M0149. The overlap in Paschen lines between the 2M1028 flare and \replaced{Flare I}{the flare discussed here} is too small, and is not included here. We note that Pa 6 ($\gamma$) in vB 10 lies below the reference line (Figure \ref{fig:PaschenDec}), showing evidence of saturation.

\subsection{He I triplet}\label{sec:hetriplet}

\begin{figure}[!t] 
\centering
\includegraphics[width=\columnwidth]{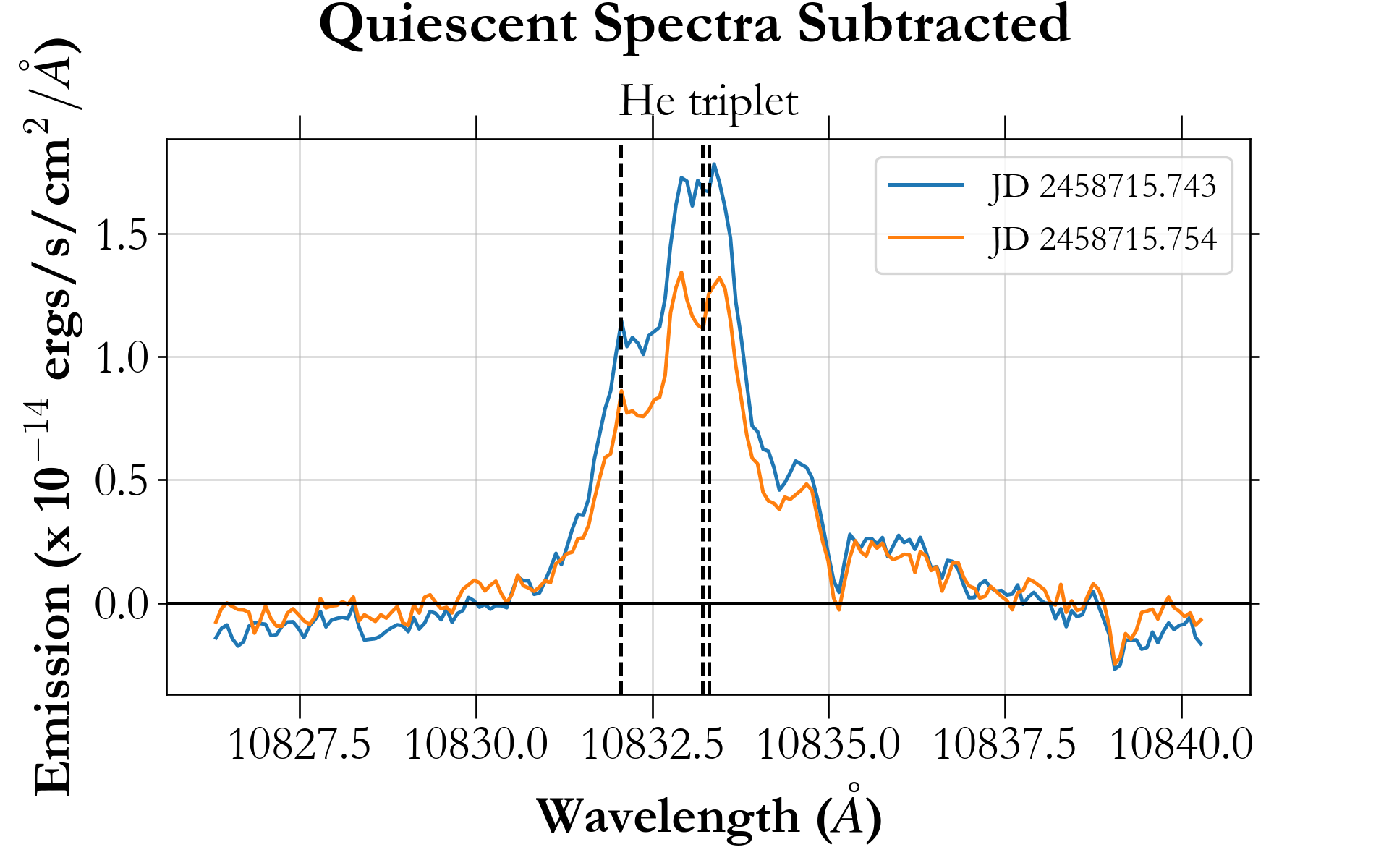}
\caption{Quiescent subtracted spectrum from HPF showing the He triplet lines seen in emission in vB 10\deleted{ during flare I}. The dashed lines mark the rest position of each line. The two different colours indicate the successive HPF observations. The triplet is one of the strongest emission features we see in the HPF spectra, and is consistent with being optically thick.} \label{fig:He_flare1}
\end{figure}

The He 10830 \AA~ triplet is one of the strongest emission features we see in the flare spectrum (\autoref{fig:He_flare1}). \replaced{For Flare I, we}{We} note that the ratio of T2/T1 for the triplet is similar to the other emission lines we observe (\autoref{fig:Lineevolution}). We note that unlike the gradual He I line decay reported by \cite{schmidt_probing_2012} for a flare around EV Lac (Figure 4), our data indicates a relatively rapid decay in the He fluxes. The two red triplet lines at 10833.2 \AA~ (Vacuum; Air = 10830.3 \AA) are blended, while the weaker bluer line is clearly distinguished (\autoref{fig:HeLorentzian}). In the optically thin regime, the ratio of I$_{\rm{blue}}$/I$_{\rm{red}}$ is $\sim 0.12$\footnote{The line strengths are proportional to the product of the Einstein coefficient $A_{ji}$, and the degeneracy $g_j$. Therefore, $I_{\rm{blue}}/I_{\rm{red}}$ $\sim$ ($A_{\rm{blue}} \cdot g_{\rm{blue}}$) / ($A_{\rm{red1}} \cdot g_{\rm{red1}}$ + $A_{\rm{red1}} \cdot g_{\rm{red2}}$) $\sim$ 0.12 \citep{kramida_notitle_2020}}. We perform a three component Voigt fit to the He I emission using \texttt{astropy} \citep{astropy_collaboration_astropy_2018}; in addition the model includes a Gaussian component to fit the red excess discussed in the next section. With the two red components of the triplet separated by just 0.1 \AA, there exists a degeneracy in the estimation of their widths and amplitudes. We obtained the best fit to the Helium excess when the two red components were bound to have the same amplitude and widths. Comparing the integrated area under each line, we obtain a ratio for I$_{\rm{blue}}$/I$_{\rm{red}}$ of $\sim 0.35$, which is consistent with being optically thick.

\cite{fuhrmeister_carmenes_2019} and \cite{fuhrmeister_carmenes_2020} report on the extensive He 10830 \AA~data set from the CARMENES survey, where the average of vB 10 observations are reported. In the 2020 paper, they explore the He 10830 \AA~ variability and discuss some cases of flaring, but not specifically for ultracool dwarfs.

\subsection{Other lines seen in emission}

In addition to the more familiar activity sensitive features discussed above, we also observe other atomic features listed in Table 1, that we could reliably measure fluxes for. The two Fe I multiplet lines at 8691 and 8826 \AA~ were also observed by \cite{fuhrmeister_multiwavelength_2008} in CN Leo, and were $\sim 10\times$ stronger. The Mg I line at 8809 \AA~ was also observed to be $\sim 30\times$ stronger in the CN Leo flare. The strength difference between the two flares is not surprising, since the CN Leo observation were close to the flare peak, and as mentioned before, the vB 10 data \replaced{for Flare I}{is} in the decay phase of the flare.

\begin{figure*}[!t] 
\centering
\includegraphics[width=\textwidth]{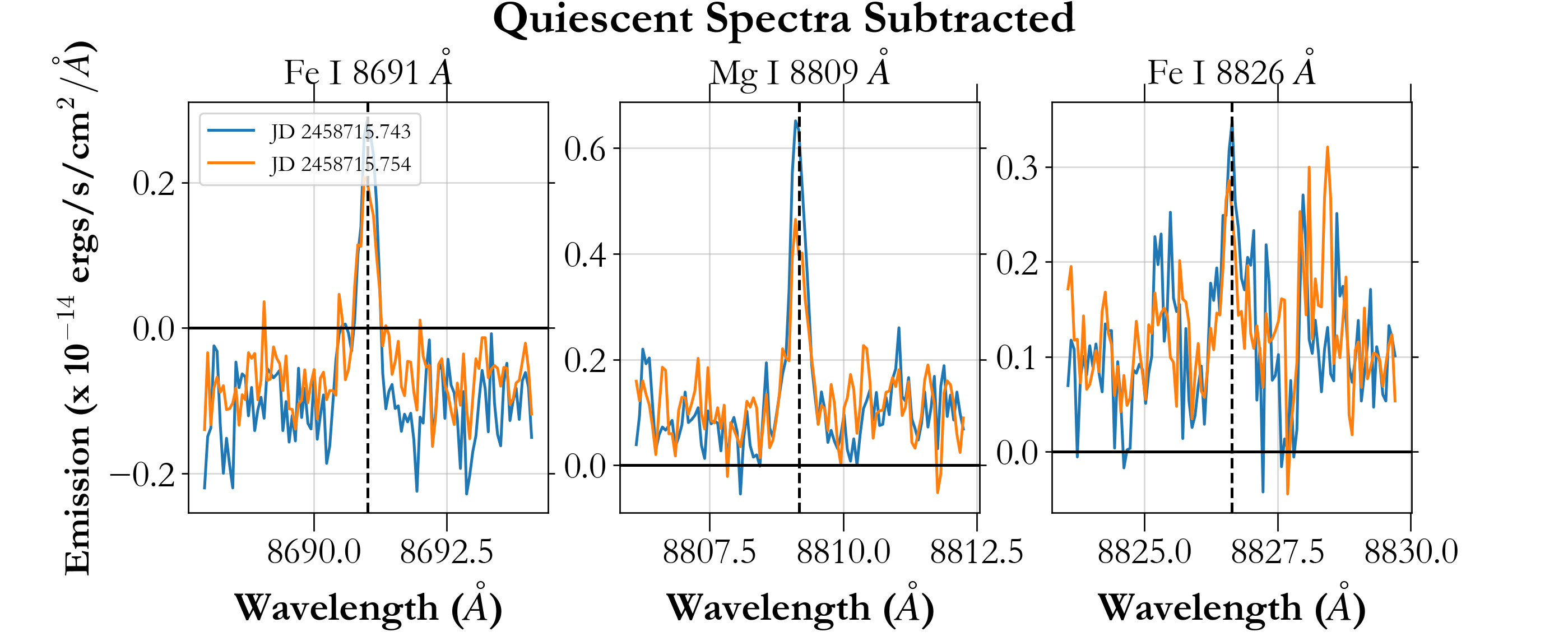}
\caption{Quiescent subtracted spectrum from HPF showing the metal lines seen in emission in vB 10 \deleted{during Flare I}. The dashed lines mark the rest position of each line. The blue and orange lines indicate T1 and T2 respectively.} \label{fig:Metal_flare1}
\end{figure*}

\begin{figure*}[!t] 
\centering
\includegraphics[width=\textwidth]{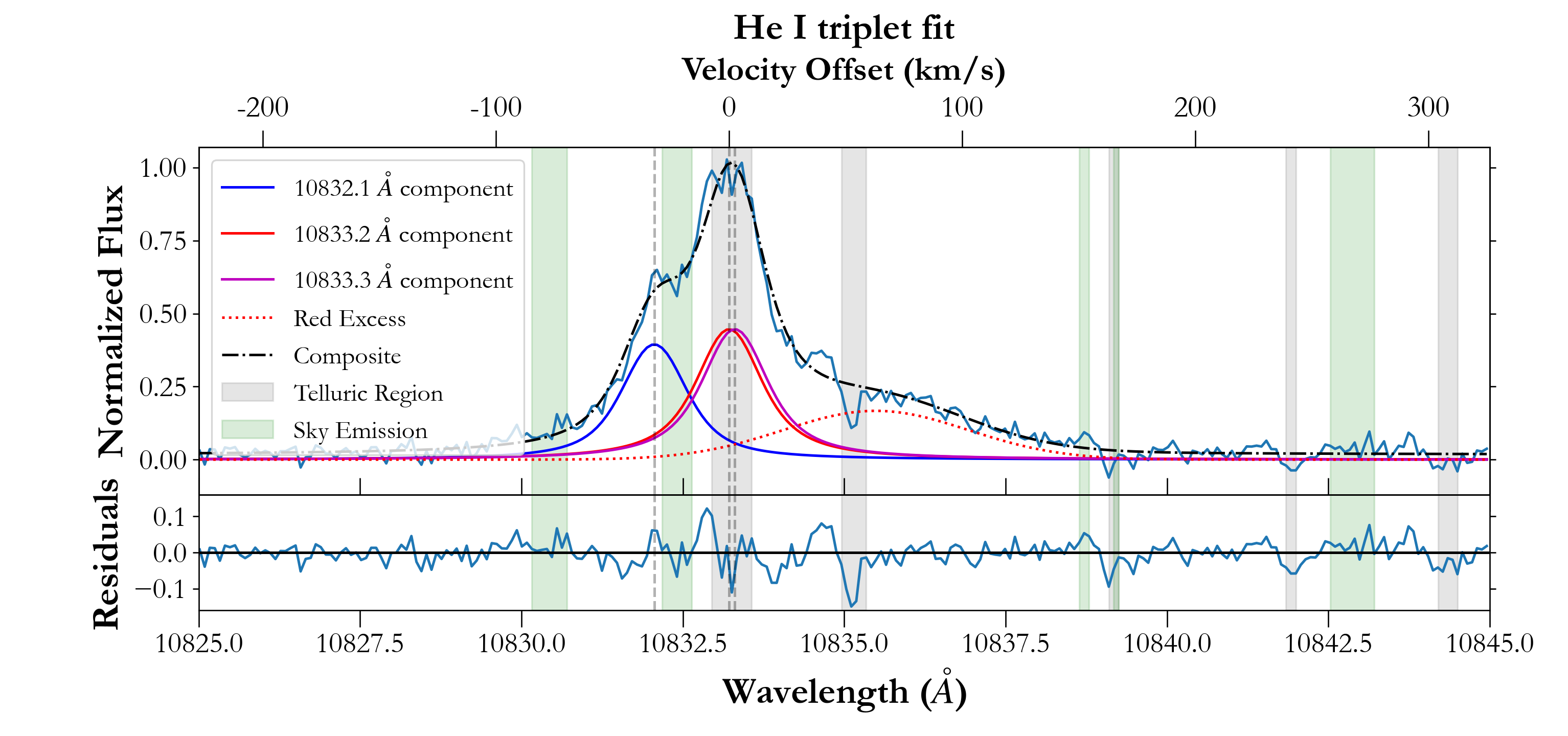}
\caption{Voigt fit to the He triplet emission for T1 (\autoref{fig:He_flare1}), showing the three components, as well as excess emission at the red wing of the triplet ($\sim 10836$ \AA). \added{The rest wavelength for each line is marked with a dashed line.} The shaded grey regions indicate where the telluric model (Section \ref{sec:telluric}) shows $> 10\%$ telluric absorption (before correction) \added{, and are masked while fitting the line profiles.} \replaced{, whereas}{Similarly,} the green regions mark the sky emission lines. We also fit the red excess with a Gaussian and find the peak of the red components at about ($10835.5$ \AA). The red excess seen here is about 30$\%$ in magnitude to the total He I triplet emission.} \label{fig:HeLorentzian}
\end{figure*}

\subsection{Red Asymmetry}\label{sec:he_asymm}
\explain{Restructured into separate section}
We note the excess emission in the red wing of the He I triplet at $\sim 10836$ \AA (\autoref{fig:HeLorentzian}), which can be attributed to downward mass motion, due to either chromospheric condensation \citep{ichimoto_h-alpha_1984, canfield_h-alpha_1990, graham_temporal_2015, graham_spectral_2020} or coronal rain \citep{antolin_observing_2012, judge_helium_2015, lacatus_explanation_2017, ruan_when_2021} depending on the temporal morphology of the emission. Coronal rain is typically seen in the gradual phase of the flare (similar to our T1-T2 observations). Conversely, chromospheric condensation is generally observed in the impulsive phase, even though there have been solar observations that show that this is not always the case \cite{graham_spectral_2020}. Given that we do not see a similar asymmetry in the Ca IRT, we infer this emission to be originating from the upper chromosphere or corona.

We fit the red excess with a Gaussian, and estimate the velocity offset for this feature (to the primary lines) \deleted{during Flare I} to be $\sim 70$ \kms{} (\autoref{fig:HeLorentzian}), which is consistent with the average of 60 -- 70 \kms{} measured by \cite{antolin_observing_2012, lacatus_explanation_2017} and $\sim 100$ \kms{} by \cite{fuhrmeister_carmenes_2018}. Using this fit we estimate the standard deviation of the Gaussian to be 1.4 \AA~(corresponding to a kT temperature of 22,000 K), and the flux emitted in this red excess to be $\sim 30 \%$ of the He I triplet. We also perform a similar analysis for the 2nd HPF visit on this flare -- T2, and obtain a similar velocity offset, width, and flux ratio. We do not measure an appreciable evolution in the velocity of this asymmetry in the higher time resolution (albeit lower S/N) SUTR time series, which is also consistent with the coronal rain hypothesis \citep{graham_spectral_2020}. \added{This also agrees with solar coronal rain measurements in post-flare loops, which follow 40 -- 80 minutes after the impulsive phase in the form of sudden condensation events  \citep[Figures 1 and 2 ;][]{ruan_when_2021}. Simulations for the Sun indicate the occurrence of quasi-periodic pulsations (QPPs) in plasma after such coronal rain events \citep{ruan_when_2021}, which have been observed around M dwarfs, most recently using photometry \citep{ramsay_tess_2021} from the Transiting Exoplanet Survey Satellite \cite[TESS;][]{ricker_transiting_2014}.}

\begin{figure}[!t] 
\centering
\includegraphics[width=\columnwidth]{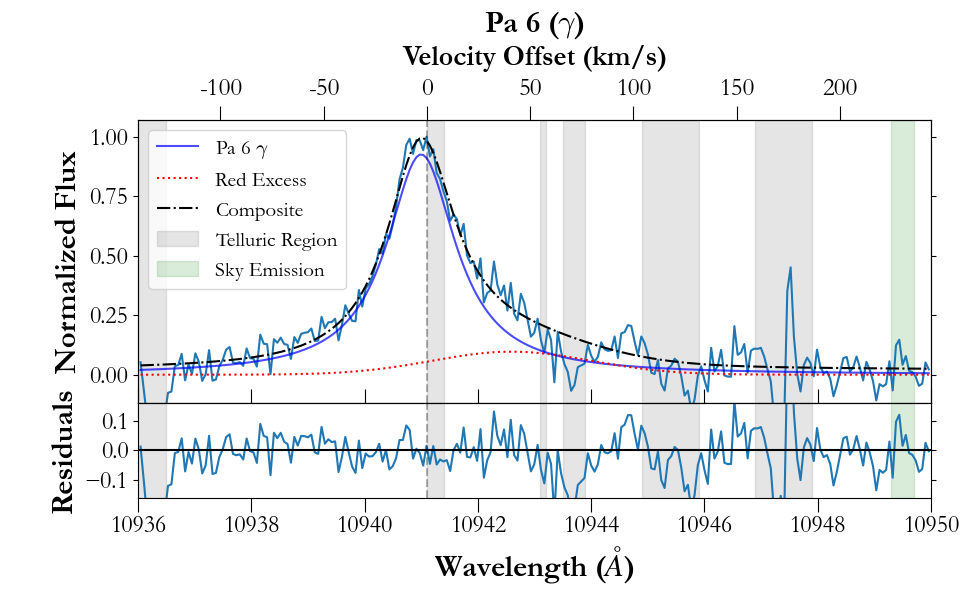}
\caption{Similar to the fit for the  He I triplet, we fit a Voigt profile to the Pa 6 ($\gamma$) line and note a very weak excess over the red wing of the Pa line. The rest velocity for the Pa line is marked by the dashed line. The shaded grey regions indicate regions where the telluric model (Section \ref{sec:telluric}) shows $> 10\%$ telluric absorption (before correction) and is masked during line fitting, whereas the green regions mark the sky emission lines. Note that the wavelength limits in this plot are larger than \autoref{fig:Pa_flare1}, to illustrate the the wings of the line.} \label{fig:Pa_red}
\end{figure}

 \added{We also note a possible weak excess redwards ($\sim$ 50 -- 150 \kms{}) of Pa 6 ($\gamma$) (\autoref{fig:Pa_red}) that is not detected for the other Pa lines. While further inference is impeded by artefacts from telluric correction, we fit a Gaussian of width similar to that of the He triplet (1.4 \AA) to this asymmetry in the Pa line, and note a velocity offset of $\sim$ 50 \kms{}, with the flux emitted in the red excess to be $\sim$ 15\% that of the Pa line.}

\subsection{Temporal evolution of atomic lines}\label{sec:linedecay}

\begin{figure}[!t] 
\centering
\includegraphics[width=\columnwidth]{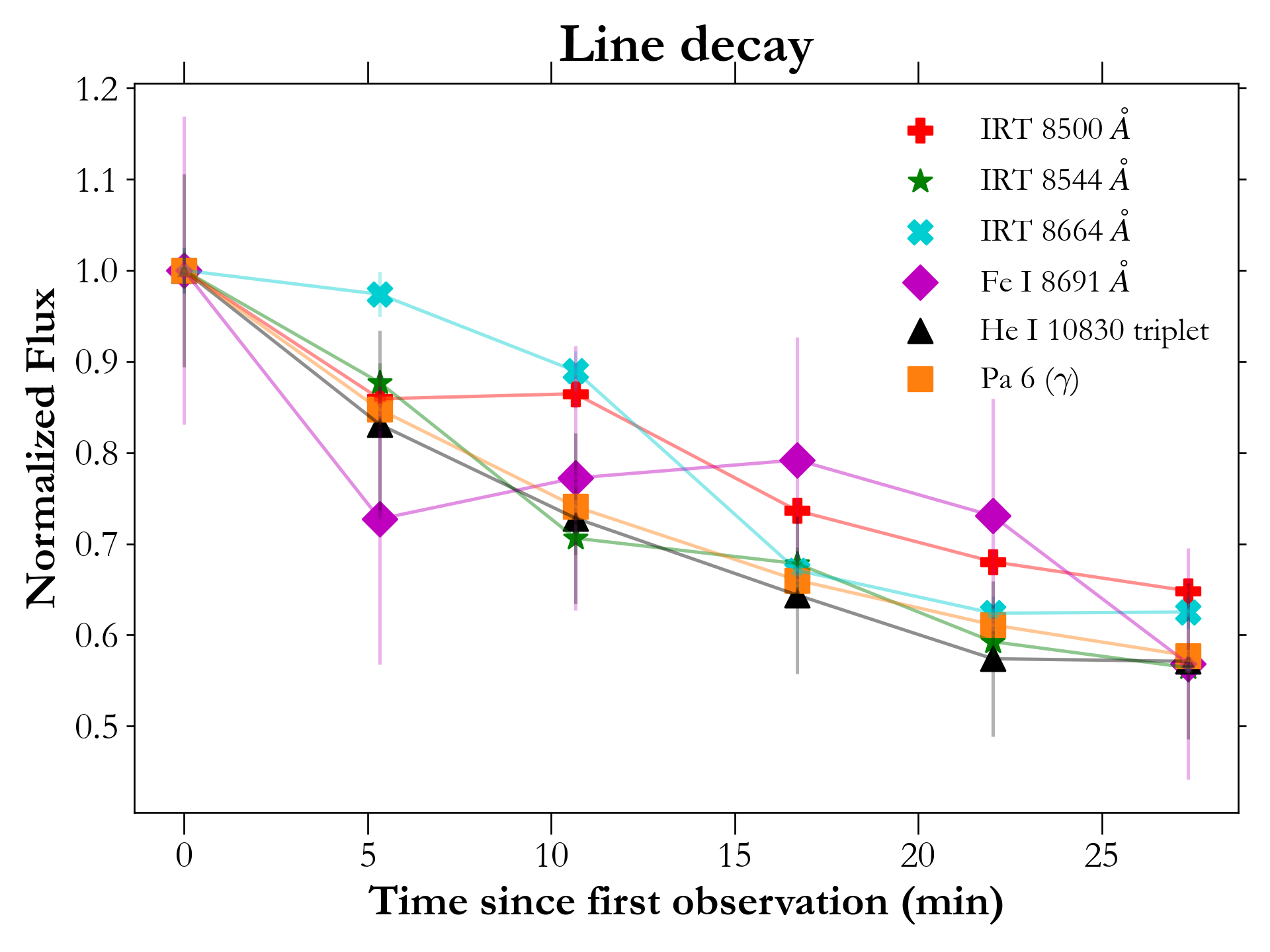}
\caption{We split the 945 s  HPF NDR exposure into 3 sub-exposures of 315 s each. For each of the lines, we show the normalized fluxes for the 6 HPF exposures \deleted{during Flare I} as a function of exposure midpoint. We fit a linear decay to the 315 s fluxes, the average slope of which is about -1.5\% min$^{-1}$. The fluxes for the NDR exposures are calculated by summing across a 2 \AA~window for the Calcium and Iron lines, and a 5 \AA~window for the Pashen lines and Helium triplet. The window is narrow enough that relative change in the instrument response across that window is negligible, and hence the errorbars depicted in this plot are based on the propagated variance for each line.} \label{fig:Lineevolution}
\end{figure}

\replaced{The Flare I}{These} observations span the gradual decay phase when the lines are reducing in strength. \autoref{fig:Lineevolution} shows the decay for the Ca IRT, an iron line, the He triplet, as well as the Pa 6 line, and can be compared to Figure 4 from \cite{schmidt_probing_2012}. In particular we note that while the He triplet fluxes decay slower than the other lines in their flare spectra of EV Lac, our analysis for vB 10 shows that this is not always true, \added{with \autoref{fig:Lineevolution} showing that all the lines analyzed here have similar rates of decay.}

\section{Discussion}\label{sec:discussion}

\subsection{Estimating \halpha{} fluxes}\label{sec:halphaestimates}

\subsubsection{Using the Ca II IRT to estimate \halpha{}}\label{sec:halpha_irt}
The \halpha{} transition is one of the most common diagnostics for flares in M dwarfs\footnote{For M dwarfs, measurements of the  Ca II H$\&$K lines are limited by the faintness of the nearby pseudocontinuum.}. The literature on flare spectra in M dwarfs is extensive but heavily weighted towards earlier spectral sub-types and visible wavelengths. While that for the ultracool M dwarfs is even more sparse --- \cite{berger_simultaneous_2008-1} summarize extensive optical data in the region 3840 -- 6680 \AA~ at $\sim 5.5$ \AA~ spectral resolution, as part of a multi-wavelength study of activity in very cool M dwarfs. The 2M1028 and 2M0149 flare observations are unique for late M dwarfs because they both have Ca II IRT and \halpha{} data simultaneously during a flare event. Since the HPF bandpass does not include \halpha{}, we use these observations to obtain a relation between \halpha{} and IRT fluxes for similar stars during flares, and by doing so, seek to compare the HPF observations of the vB 10 flare with other similar events.

We do not use the L$_{\rm{line}}$ / L$_{\rm{bol}}$ values reported with these flare observations, since the luminosity estimates are based on obsolete parallax measurements. We instead use \gaia{} EDR3 parallax measurements (\autoref{tab:parallaxes}) for these three targets (vB 10, 2M1028 and 2M0149).

To get a scaling of F$_{\rm{IRT}}$ to F$_{\rm{H}\alpha}$ for 2M0149 we take the ratio of \halpha{} fluxes to the sum of the three IRT fluxes. The average of this scaling factors for the four observations from \cite{liebert_2mass_1999} Table 1 is 6.68. Following the same procedure for 2M1028 yields a factor of 6.64. We use an average of the scaling factor from 2M0149 and 2M1028, i.e., 6.66.

To estimate the \halpha{} flux \deleted{during Flare I}, we add the three IRT fluxes for T1 to obtain a total of 8.34 $\times 10^{-14}$ \flux{}, multiplying which by 6.66 gives a vB 10 \halpha{} estimate of 56 $\times 10^{-14}$ \flux{}. The T2 fluxes are about 75$\%$ of the first spectrum. These extrapolated \halpha{} fluxes \deleted{for Flare I} are comparable to the fourth and weakest observation reported by \cite{liebert_2mass_1999} for 2M0149, and about 2$\%$ of the values reported by \cite{schmidt_activity_2007} for 2M1028.  Conversely, \halpha{} fluxes reported by \cite{berger_simultaneous_2008-1} are about 3--4$\times$ weaker than \replaced{the}{our} estimates \replaced{of vB 10 Flare I}{for vB 10}.

\subsubsection{Using the Pa lines to estimate \halpha{}}\label{sec:halpha_paschen}
We also use the Paschen lines to estimate the \halpha{} flux for \replaced{Flare I}{vB 10}. We continuum subtract the 2M0149 spectrum \citep[][Table 1]{liebert_2mass_1999}, and average the ratio of \halpha{} to Pa 7 ($\delta$) across their four observations to obtain a ratio of 19. Multiplying this ratio to the Pa 7 ($\delta$) flux from \autoref{tab:line_flare1} - T1, estimates the \halpha{} flux to be 25 $\times 10^{-14}$ \flux{}, which is $\sim 2\times$ lower than that obtained from the IRT analysis.

\subsection{Energy released during the flare}\label{sec:flareenergy}

\begin{figure}[!t] 
\centering
\includegraphics[width=\columnwidth]{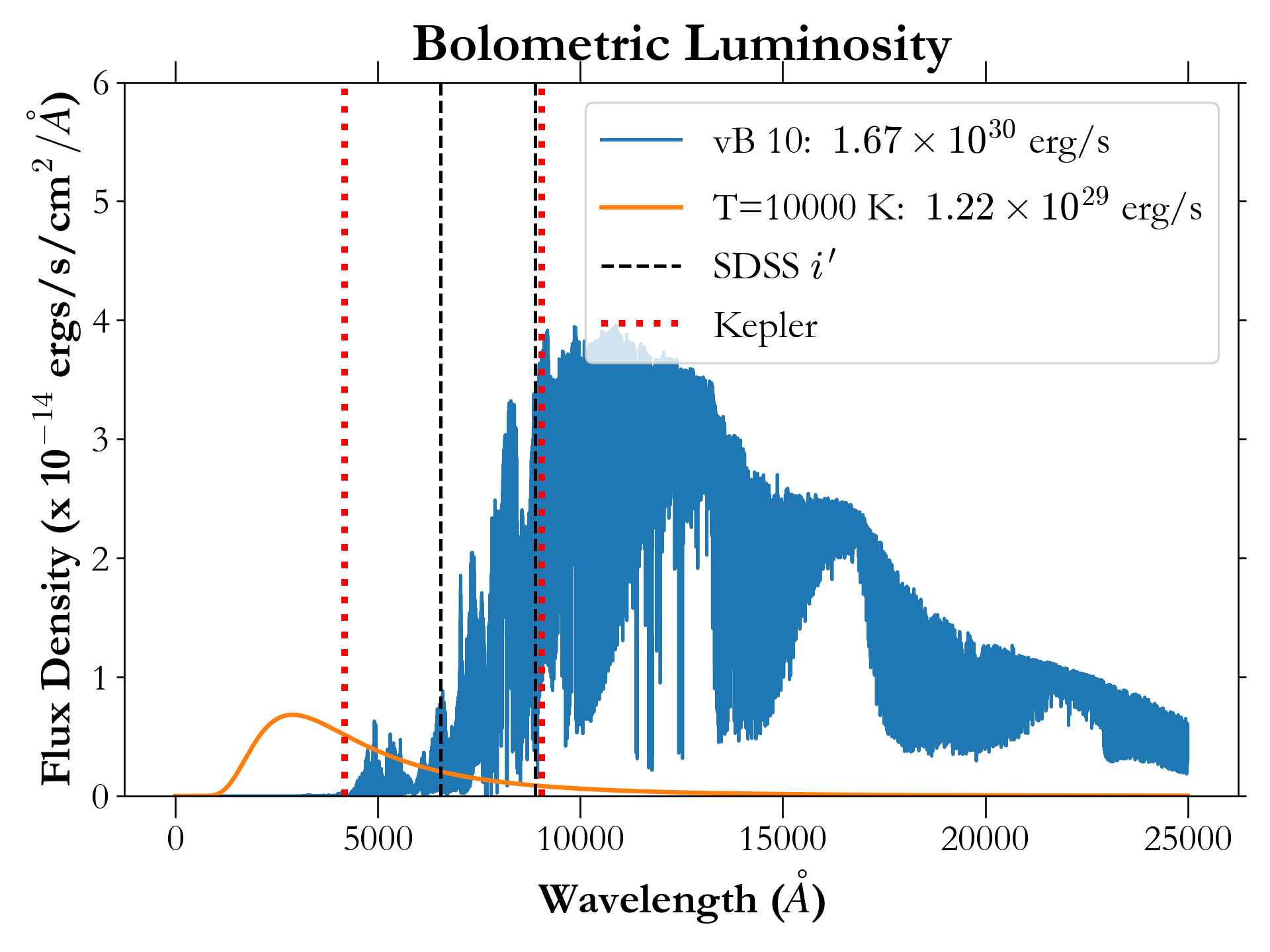}
\caption{In this plot we show a \texttt{BT-Settl} model spectrum \citep{allard_models_2012} at 2700 K for vB 10, in addition to a scaled 10,000 K blackbody. The black dashed line indicates the SDSS $i^\prime$ bandpass used for the observations in Section \ref{sec:het_acam} to estimate the continuum enhancement, meanwhile the red dotted line shows the \textit{Kepler} bandpass used for the analysis in Section \ref{sec:flarefrequency}. The bolometric luminosity for vB 10 is 1.67 $\times 10^{30}$ erg/s \citep{tinney_faintest_1993}, while that for the 10,000 K blackbody is 1.22 $\times 10^{29}$ erg/s. This suggests that right before the HPF observations the bolometric luminosity of the flare was $\sim 7 \%$ of the total luminosity of the star.} \label{fig:continuumenhancement}
\end{figure}

\begin{figure*}[!t]
\gridline{\fig{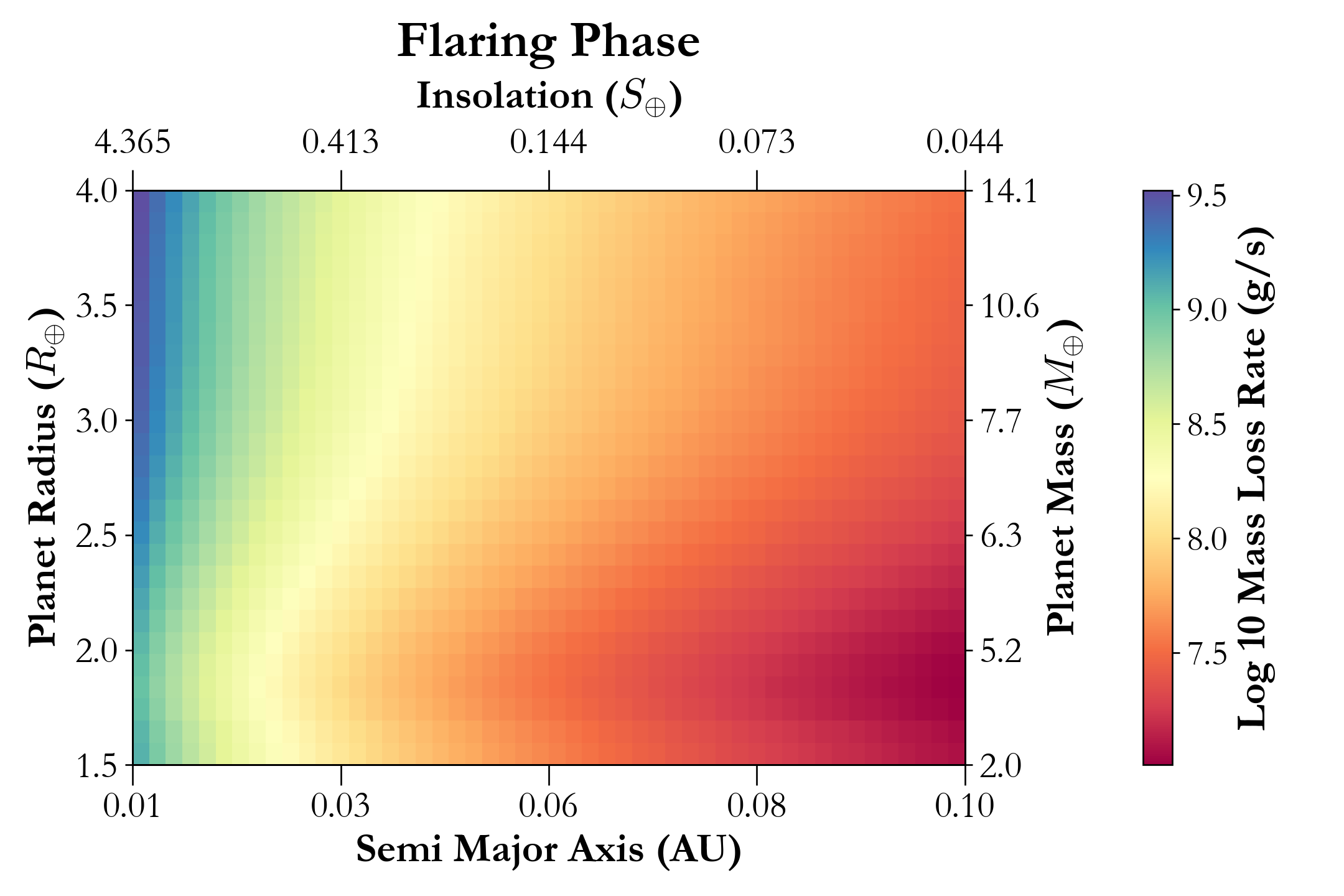}{0.48\textwidth}{{\small a) Mass loss rate for flaring phase}}    
          \fig{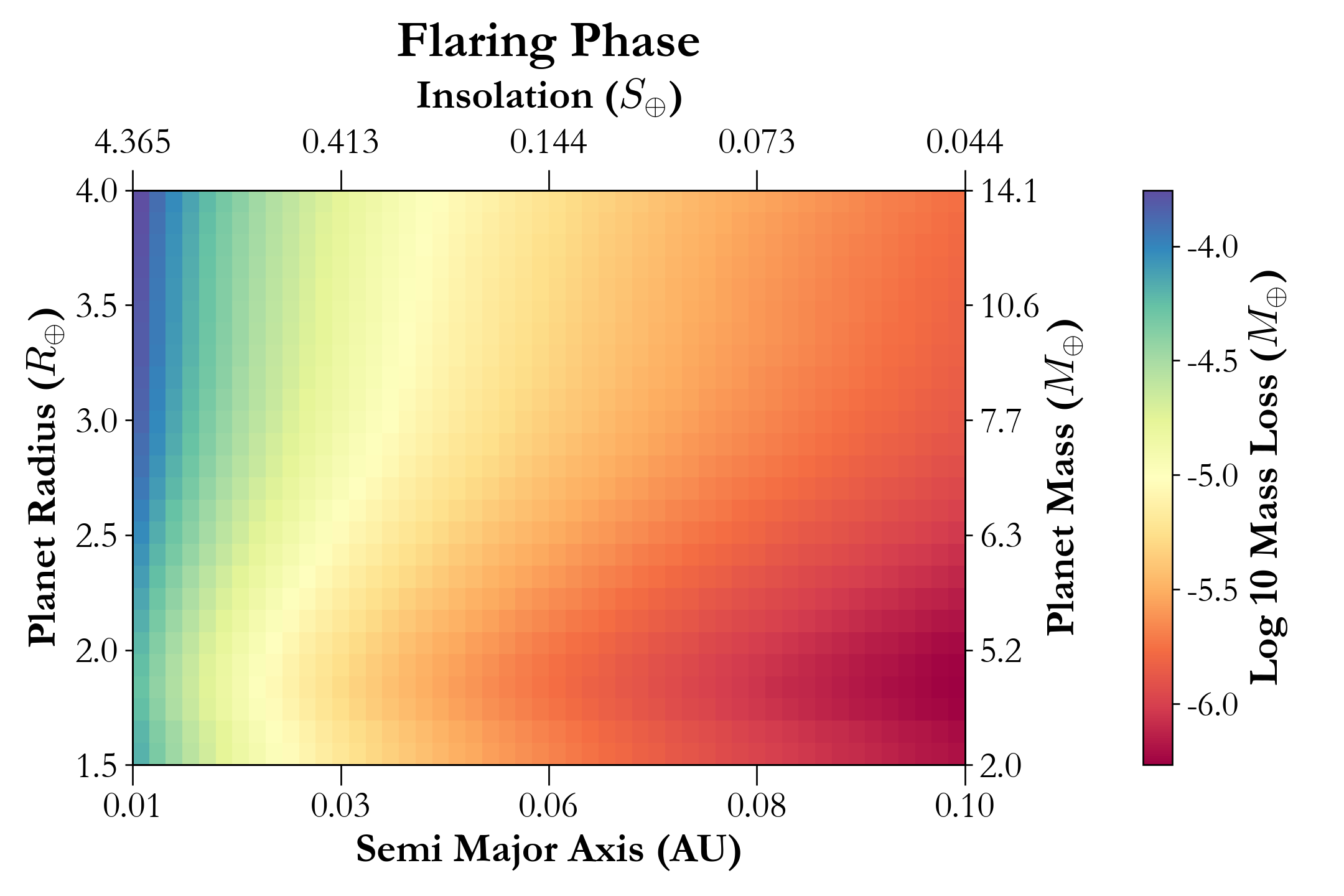}{0.48\textwidth}{ \small b) Total mass loss for flaring phase}} 
\caption{\small The atmospheric mass loss across a range of small planet radii (1.5 -- 4 \earthradius{}), and semi-major axis spanning 0.01 -- 0.1 AU using \autoref{eq:massloss}. \textbf{a)} This panel shows the rate of mass loss, assuming a a high energy luminosity  ($L_{HE} = 10^{27}$ erg/s) equivalent to \cite{fleming_x-ray_2000}.  To calculate the insolation at difference separations for vB 10, we use the bolometric luminosity of log L$_{\rm{bol}}$ = -3.36 \citep{tinney_faintest_1993}} \textbf{b)} We propagate the mass loss rate over 1 Gyr to calculate the cumulative atmospheric mass loss for a small planet around vB 10, with flare duration of 1 hour and frequency of once per 100 hours, as estimated in Section \ref{sec:flarefrequency}. As discussed in Section \ref{sec:photoevaporation}, the mass loss during the quiescent phases should be approximately double the loss from such flares. The relatively low mass loss rate estimated for vB 10 is consistent with its low \vsini{}, and is likely to have been higher earlier in its lifetime.\label{fig:FlaringMassLoss}
\end{figure*}

We follow a procedure similar to \cite{gizis_k2_2017}, to estimate the bolometric energy released during the decay phase of this flare from the HET ACAM photometry. We use \texttt{BT Settl} model spectrum \citep{allard_models_2012} at 2700 K \citep[vB 10 \teff{} = 2745 K;][]{fouque_spirou_2018}, which is scaled to the 2MASS \citep{cutri_2mass_2003} J mag = 9.9 (Section \ref{sec:absolute}). To simulate the flare spectra, we use a blackbody at 10,000 K \citep{kowalski_new_2015}, generated with \texttt{astropy} \citep{robitaille_astropy_2013, astropy_collaboration_astropy_2018}\footnote{We note that the 10,000 K blackbody used is a simplified model. \cite{kowalski_time-resolved_2013} have shown continuum enhancements consistent with blackbody of temperatures ranging from 9000 to 14,000 K for a sample of mid-type M dwarf flares. While \cite{gizis_kepler_2013} observe a flare around an L1 dwarf, and report a white light flare continuum that corresponds to a 8000 K blackbody.}. This is then scaled to match the continuum enhancement of $\sim 12\%$ seen in SDSS $i^\prime$ using the HET ACAM photometry before T1 (Section \ref{sec:het_acam}). These spectrum are plotted in \autoref{fig:continuumenhancement}, using which we find the bolometric luminosity during this phase to be 1.22 $\times 10^{29}$ erg/s, which is about 7$\%$ the bolometric luminosity of the vB 10.

We refer to the analysis of K2 light curves for flares around ultracool dwarfs by \cite{paudel_k2_2018} to approximate the total energy released during this flare. They analyze the temporal morphology of white light flares around 2M0835+1029 (an M7 dwarf), and 2M1232-0951 (an L0 dwarf), as observed by K2 in the \textit{Kepler} bandpass\footnote{Spanning 4183 -- 9050 \AA~with a FWHM of 3993 \AA.}. We compare the rate of decay in the IRT fluxes (Section \ref{sec:linedecay})\footnote{This is consistent with the slope to a linear fit to the Pa 6 ($\gamma$) flux decay.}, with the K2 flare light curve for 2MASS J08352366+1029318 and 2MASS J12321827-0951502. These two flares reach a comparable slope, when the fluxes reach 12\% and 8\% of the peak flux, respectively. Taking the average of these two, we estimate that our measurements of the bolometric luminosity using the HET ACAM photometry are at about 10$\%$ of the peak emission of 10$\times$ $1.22 \times 10^{29}$ erg/s or $\sim 1.22 \times 10^{30}$ erg/s. This is $\sim 70 \%$ of the total bolometric luminosity of the star.  Furthermore, we integrate the area under the two K2 flare curves and obtain the total energy released during the flares to be $\sim 8.5\times$ the peak energy, i.e., for the vB 10 flare, the total energy should approximate $\sim 10^{31}$ erg. We note that this estimate is an approximate, with an error of 50 -- 100 \% because of the assumptions made regarding the temporal morphology of these flares, as well as the error present in the flux estimates from the HET photometry.  Since we do not have observations spanning the entirety --- both temporally and in wavelength --- of the flare, we use these comparisons to understand the nature of the flare.

\subsection{Estimating the flare frequency}\label{sec:flarefrequency}
\cite{paudel_k2_2018} average the flare rates for ultracool dwarfs across different spectral types, as a function of the total energy released in the \textit{Kepler} bandpass during the flare. To estimate the flare frequency using these, we need to calculate the energy released \deleted{during Flare I} for vB 10. To do so, we multiply a blackbody of 10,000 K (Section \ref{sec:flareenergy}) with the \textit{Kepler} bandpass (\autoref{fig:continuumenhancement}), and integrate it to estimate the luminosity during the decay phase in the \textit{Kepler} band to be $2.5 \times 10^{28}$ erg/s. Based on the scaling obtained in the previous section, this corresponds to $2 \times 10^{30}$ erg  released in the \textit{Kepler} bandpass during the flare. In Figure 10 and 12a from \cite{paudel_k2_2018}, this flare energy corresponds to an average flare frequency of once every 100 or 120 hours respectively.

We also use HPF observations to place an independent upper limit on the cadence of flares of similar energy on vB 10. We attempted to bound the flare rate of vB 10 under the assumption that---as shown for other M dwarfs \citep[e.g.][]{pettersen_spectroscopic_1989, chang_photometric_2015, li_waiting_2018}---the star flares at times well described by a Poisson distribution.  With only one flare in the data set, we are unable to fully constrain the flaring frequency of VB 10 for such flares, but we can provide an upper bound.

We created simulated time series of flares, representing the flare series as a square wave with Poisson-distributed pulses.  The height of each pulse is constant, as we seek to constrain the frequency of any flares energetic enough to be clearly noticed in the HPF spectral series.  The temporal spacing of the simulated flares is determined by a wait time (the average time between flare events), and the number of flares per flare series is set such that the time baseline of the series matches or exceeds the baseline of the HPF data. We run this simulation for a flare duration (=pulse width) of 1 hour.

For each flare wait time, 10,000 trials are performed, with each trial creating a new flare series and new flare distribution. For each trial, we consider an individual simulated flare to be ``recovered" if it occurs during a time matching a timestamp from our HPF time series.  After all 10,000 trials, the numbers of “recovered” flares are then organized into a histogram to determine the peak number of recovered flares for a given wait time. To avoid double counting where two data points may occur within one flare, a flare is no longer counted in that series/trial once one data point is found to occur within the flare. In this case, we looked for a peak of one recovered flare, just before the peak transitions to two. At this point, the wait time is the maximum flare frequency (or in other words, the star cannot flare more often than wait time days). Therefore, for a flare duration of 1 hour \citep{paudel_k2_2018}, we obtain a maximum flare frequency of once every 2.5 days (or 60 hours). This upper limit to the frequency is consistent with the estimate of $\sim 100$ hours based on the scaling from the K2 ultracool dwarf flare observations \citep{paudel_k2_2018}.


\subsection{Implication for Photoevaporation}\label{sec:photoevaporation}

\deleted{
The He 10830 \AA~emission line formation can have components of photo-ionization, recombination, and collisional excitation, with the latter becoming more important in the denser atmospheres of M dwarfs \citep{sanz-forcada_active_2008}. The X-ray and Extreme Ultra-violet (EUV) emission seen in vB 10 flares \citep{linsky_stellar_1995, berger_simultaneous_2008-1} indicates the presence of hot plasma required for the excitation of Helium. Indeed, \cite{schmidt_probing_2012} noted that their model underestimates the He 10830 \AA~emission, compared to their observations of the flare on EV Lac. \cite{allred_unified_2015} discuss a unified approach to solar and stellar flares that features electron beams using an algorithm, that was then further developed by \cite{kerr_crucial_2021}. The collisional and photo-ionization of He I required to produce the observed emission requires a EUV rich environment. The strength of He 10830 \AA~emission we observe in the later stages of \replaced{Flare I}{the vB 10 flare}, suggests an EUV and X-ray (XUV) environment similar to that observed by \cite{linsky_stellar_1995, berger_simultaneous_2008-1}.}

Atmospheres of close in exoplanets are susceptible to photoevaporation due to high energy ionizing radiation \citep{owen_planetary_2012}. Similar EUV and X-ray radiation has been reported for vB 10, during both quiescent \citep{fleming_quiescent_2003} and flaring phases \citep{linsky_stellar_1995, fleming_x-ray_2000, berger_simultaneous_2008-1}.  We attempt to place limits on the mass loss rates of a hypothetical small planet ($R_p < 4$ \earthradius{}) across a range of radii and separation from the star, using a simplified version of the formalism from \cite{erkaev_roche_2007}, \cite{penz_x-ray_2008}, and \cite{owen_planetary_2012} to estimate the mass loss rate ($\dot{M}$):

\begin{equation}\label{eq:massloss}
    \dot{M} = \frac{\pi R_p^3}{GM_p} \frac{L_{HE}}{4\pi a^2},
\end{equation}

where $R_p$, and $M_p$ are the planetary radius and mass respectively, $a$ represents the semi-major axis, $G$ the gravitational constant, and $L_{HE}$ the high energy ionizing radiation luminosity. We use the \texttt{MRExo} \citep{kanodia_mass-radius_2019} mass-radius relation for M dwarf planets, to predict planetary masses for a range of radii.   \replaced{For our Flare I observations, we do not have contemporaneous coverage of the EUV and X-ray. However, the He 10830 \AA~ emission suggests an XUV radiation rich environment accompanying the flare.}{We are unable to constrain the Extreme Ultra-violet (EUV) emission for vB 10 with our data spanning the decay phase of this flare. However, we expect the early stages of this flare to be accompanied by high energy EUV and X-ray radiation based on previous flare observations of vB 10.} We calculate the mass loss rate over 1 Gyr for two cases i) using the estimates for frequencies of such flares from our observations, and an $L_{HE}$ equivalent to the X-ray flare reported by \cite{fleming_x-ray_2000} ($L_{HE} = 10^{27}$ erg/s) as an upper limit (\autoref{fig:FlaringMassLoss}), and ii) using the quiescent X-ray luminosity of $2 \times 10^{25}$ erg/s, as reported by \cite{fleming_quiescent_2003, berger_simultaneous_2008-1}. In the first case, we use the flare frequency of once every 100 hours (Section \ref{sec:flarefrequency}), and a duration of 1 hour \citep[based on][]{paudel_k2_2018}, i.e., a duty cycle of $\sim 1\%$ (this is consistent with the $3\pm1\%$ duty cycle estimated by \cite{hilton_galactic_2011} for the M7 -- M9 spectral subtype.). Conversely, in the case of quiescent high energy radiation, the X-ray luminosity is about 1/50th that of the flares. Given the 1\% duty cycle for such flares, and the characteristics stated above, the mass loss during the quiescent phase should be \replaced{2x the loss from the flaring phase.}{higher (2x) than that from the flaring phase.}

For reference, according to the planetary models from \cite{lopez_understanding_2014}, the atmospheric envelope mass fraction for sub-Neptune in the radius valley \citep{fulton_california-kepler_2017} with radius = 1.7 \earthradius{}, mass = 3.4 \earthmass{} and insolation of 1 $S_{\oplus}$ (Semi-Major Axis = 0.02 AU), should be 0.2\% or an atmospheric mass of $7 \times 10^{-3}$ \earthmass{}. Therefore, the total atmospheric mass loss over a Gyr is 0.5\% of the atmospheric mass given similar flares (\autoref{fig:FlaringMassLoss}). Using this hypothetical sub-Neptune in the \replaced{HZ}{Habitable Zone} of vB 10 as an example, we place limits on the atmospheric mass loss from flaring and quiescent high energy radiation.

We include the caveat that we have assumed the rate of radiation to be constant with stellar age in this work. vB 10 has a low \vsini{} \citep{reiners_carmenes_2018}, and is likely to have been more active with more frequent flares, and a higher L$_{\rm{X}}$/L$_{\rm{bol}}$ when it was younger \citep{paudel_k2_2018, paudel_k2_2019, mcdonald_sub-neptune_2019}, resulting in a larger mass-loss rate\footnote{Furthermore, for simplicity we do not consider the Roche-lobe effects discussed by \cite{erkaev_roche_2007} in these calculations.}. vB 10 was not observed by the \textit{Kepler} or K2 mission, but will be observed by the TESS in Cycle 4 during Sector 54 (July 2022). This will help place better limits on the flare rate as a function of energy for vB 10, similar to work from \cite{feinstein_flare_2020}, which will also be useful to constraint planetary atmospheric losses due to photoevaporation.

\section{Conclusion}\label{sec:conclusion}

In this work, we present observations of a flare around the ultracool M8 dwarf vB 10, which include high resolution NIR spectra from HPF, as well as photometry from the acquisition camera (ACAM) on HET. We draw the following conclusions based on the HPF spectra -

\begin{itemize}
\item We use the HPF spectra to measure emission in the calcium infrared triplet, Paschen lines, the He 10830 \AA~ triplet, as well as iron and magnesium lines.
\item  We present evidence for observations of coronal rain on this star, which is not only the first time it has been observed on such a late type star, but also the first time it has been observed in the He I triplet for M dwarfs. This is in the form of an asymmetry in the He I triplet, $\sim$ 70 \kms{} redwards of the primary lines, \added{as well as tentative evidence of a similar asymmetry for the Pa 6 ($\gamma$) line.}
\item We flux calibrate the HPF spectra and estimate the absolute flux emitted in the various lines during the flare. We also note that the Calcium, Paschen and Helium lines are consistent with being optically thick.
\item Utilizing the non destructive readout capability of the HPF detector we reconstruct the temporal evolution during the gradual decay phase of the flare. 
\end{itemize}

The large HET aperture ($\sim$ 10 m), allows for high resolution spectral observations of this ultracool faint dwarf, which help place strict upper limits ($|$v$|$ $<$ 5 \kms{}) on the velocity offsets of the emission features in the spectra. Detailed hydrodynamic modelling of the atmospheres of these faint ultracool dwarfs, with their molecular hazes at low \teff{} is notoriously difficult. However, ongoing RV surveys should provide for serendipitous flare observations similar to those presented here, which should help extend the models to these lower temperatures. 

We measure a continuum enhancement in the ACAM photometry, which we use to calculate the bolometric luminosity of the flare assuming a 10,000 K blackbody. We estimate that at its peak, the flare had a bolometric luminosity $\sim 70 \%$ that of the star itself! Using scaling relations developed from K2 observations of ultracool dwarfs, we also estimate the frequency of similar flares around vB 10. Finally, we discuss the XUV radiation that is released during such flares, and its impact on the atmospheric evolution of planets.

\appendix
\section{Spectral Reduction and Correction}\label{appsec:spectralreduction}


\subsection{Telluric Correction}\label{sec:telluric}
\begin{figure*}[!t] 
\centering
\includegraphics[width=\textwidth]{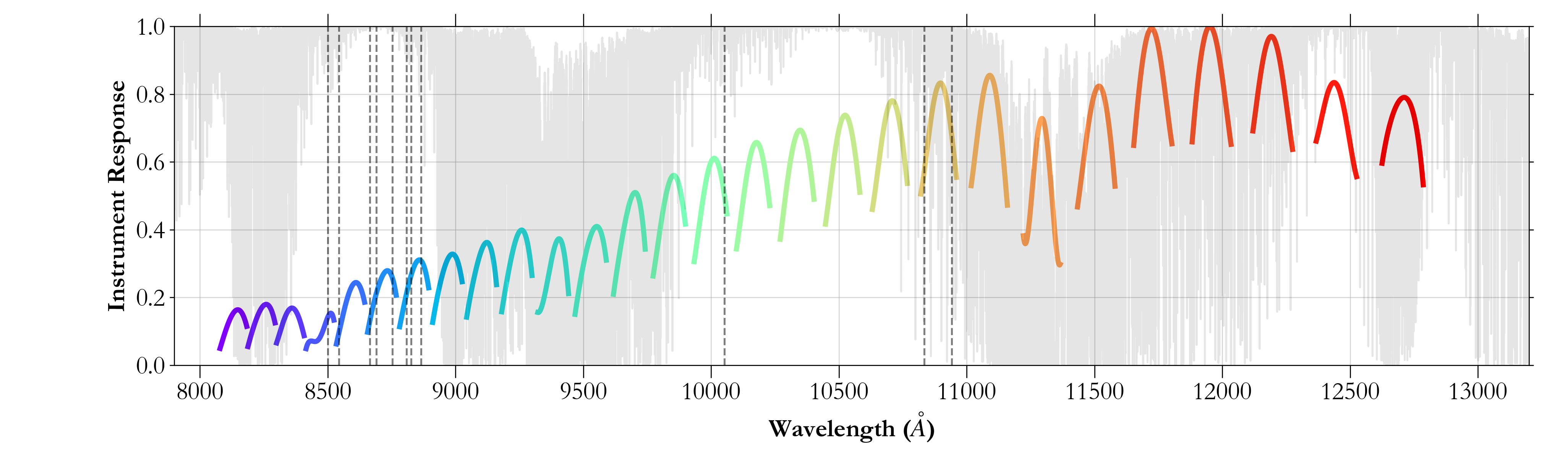}
\caption{The relative instrument response as a function of wavelength. The dashed lines mark the positions of the lines considered in this analysis (\autoref{tab:line_flare1}). Note the irregularly shaped HPF instrument response in the wavelength region spanning the first two IRT lines (Order Index 3--4; $\sim$ 8490 -- 8550 \AA), which is attributed to an absorption band in the multilayer HPF mirror coating. The shaded spectrum in the background is a representative telluric model generated from \texttt{TelFit} \citep{gullikson_correcting_2014}. This indicates the regions of heavy telluric absorption which reduce the local instrument response in particular orders.} \label{fig:InstrumentResponse}
\end{figure*}

We perform telluric correction on the sky subtracted spectrum following a procedure similar to \cite{robertson_proxima_2016}. We use the sky subtracted spectrum after correcting for the intra (and inter) order instrument response (Section \ref{sec:relative}). The telluric correction is performed using the \texttt{TERRASPEC} code \citep{bender_sdss-het_2012, lockwood_near-ir_2014}, which uses the LBLRTM radiative transfer code \citep{clough_atmospheric_2005} to generate a synthetic telluric absorption function which calculates a telluric model using an observer's altitude, target zenith angle, and telluric conditions integrated over a given science observation. LBLRTM uses the HITRAN 2012 database of molecular lines \citep{rothman_hitran2012_2013} and can access a variety of standard atmospheric models. The telluric absorption function is combined in a forward model with the instrument profile, which we parameterize as a series of overlapping Gaussian functions \citep[e.g.,][]{valenti_determining_1995, endl_planet_2000}. We use non-linear least-squares minimization \citep{markwardt_non-linear_2009} to optimize the integrated column depths of the atmospheric molecules and instrument profile parameters to best match the observed science spectrum. To process the HPF spectrum, we used the U.S. standard atmospheric model \citep{national_us_1992}, with the instrument profile as described above. We use the water-dominated telluric regions in our spectrum to get initial values for water column depth and instrument profile. We then use this as the initial values for each order-by-order fit to the HPF spectrum. We opt to fit each order separately to account for any slight changes in the instrument PSF as the wavelength increases. However, there are still wavelength regions that are quite heavily contaminated by tellurics (mainly water vapour), and are shown in \autoref{fig:InstrumentResponse} as the shaded spectrum.

In addition to the telluric absorption correction, we also perform a sky subtraction to correct for hydroxyl radical OH sky emission lines using the simultaneous sky spectrum obtained by HPF's sky fiber \citep{kanodia_overview_2018}. Correction of the sky emission features is especially important for the He 10830 \AA~ region, and is performed by scaling and subtracting the sky spectrum from the science spectrum \citep{ninan_evidence_2020}.

\subsection{Velocity Correction}\label{sec:velocity}
The telluric-corrected spectrum is shifted to the solar system barycentric frame using \texttt{barycorrpy} \citep{kanodia_python_2018}, the Python implementation of the algorithms from \cite{wright_barycentric_2014}. We estimate the absolute RV of vB 10 to be 35.5 \kms{}, which is calculated by stepping the vB 10 spectrum against HPF spectrum of GJ 699 (after shifting the GJ 699 spectrum to the stellar rest frame). in velocity space, and minimizing the $\chi^2$. This is consistent with the absolute RV from estimates from \cite{mohanty_rotation_2003}, \cite{zapatero_osorio_infrared_2009}, \cite{reiners_carmenes_2018}, and also the \gaia{} DR2 measurement for GJ 752 A\footnote{Note that there are no \gaia{} radial velocity estimates for the bulk velocity of vB 10.} (the brighter binary companion).

\begin{figure} 
\centering
\includegraphics[width=\columnwidth]{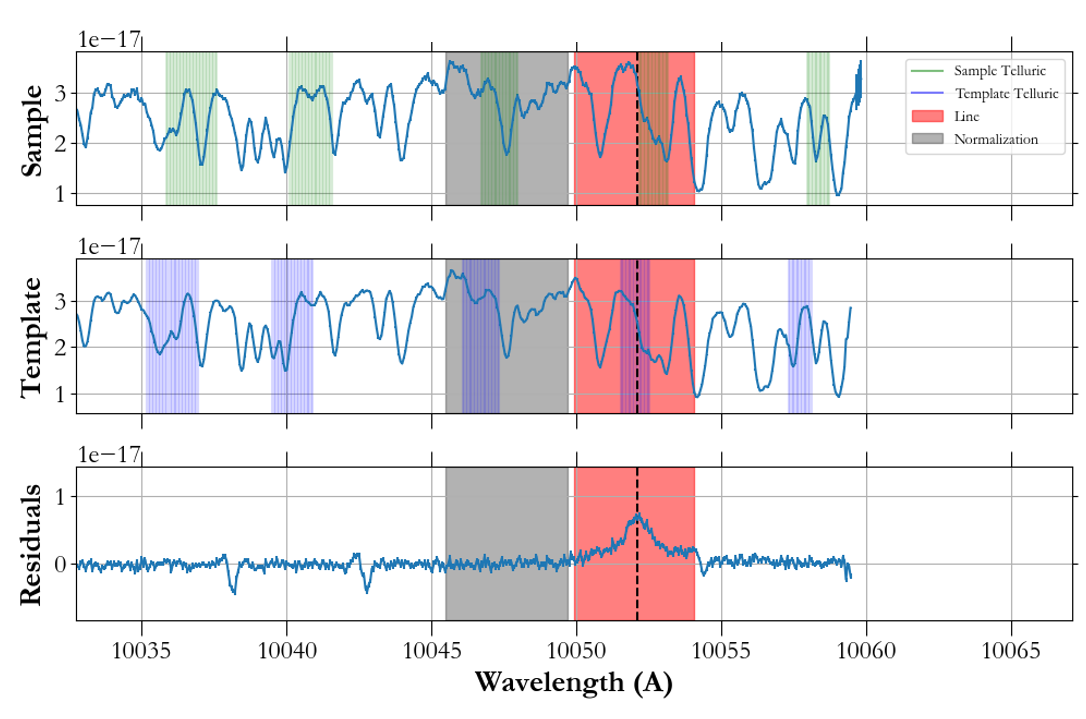}
\caption{An example of the spectral subtraction, with the sample spectrum shown in the top panel, the template in the middle, and the residuals in the bottom. While the excess at the position of the  Pa 7 ($\delta$) line is apparent in the residual spectrum, it is harder to tease out before subtracting a quiescent template. } \label{fig:SpectralSubtraction}
\end{figure}

\subsection{Relative instrument response calibration}\label{sec:relative}
To estimate the ratios of flux emitted across different lines we need to correct for the intra-order instrument response variation due to the grating blaze function, as well as the inter-order overall chromatic instrument response. In addition to the grating blaze function, for HPF order index 3 ($\sim$ 8410 -- 8530 \AA), we also see a higher order chromatic effect that is attributed to a flaw in the multilayer reflection coating in HPF (\autoref{fig:InstrumentResponse}). We also need the flux calibration to convert from measured photo-electrons to spectral flux density units (\specificflux{}).

We first modelled  HPF spectrum of the A0 star HR 3437 with a \vsini{} broadened\footnote{\vsini{} broadening of 193 \kms{} \citep{diaz_accurate_2011}} model spectrum of Vega \citep{castelli_model_1994}.  The broadened model spectrum were interpolated onto the stellar wavelength grid, and then divided from the HPF spectrum. The ratio (stellar to model) spectrum represents the relative instrument response, and was fit using 4th degree Chebyshev polynomials \citep{chebyshev_theorie_1854}. The converging Paschen series\footnote{Atomic Hydrogen transitions to $n=3$ level, with $\lambda_{n_{\infty}}$ = 8207 \AA~(Vacuum; Air = 8204 \AA)} \citep{paschen_zur_1908} of lines \citep{kramida_critical_2010, kramida_notitle_2020} in this spectrum of HR 3437 are not accurately modelled by the Vega spectrum, and hence need to be masked. Despite the masking, lines at the edges of the HPF orders skew the response fit, and prevent robust unbiased estimation of the instrument response.

We then used HPF observations of the G3 dwarf star 16 Cyg B, alongside CALSPEC model spectrum for the same target \citep{bohlin_techniques_2014, bohlin_new_2020}. The CALSPEC model spectrum does not require \vsini{} broadening, and was interpolated onto the stellar wavelength grid, and then normalized to 1. The ratio of stellar to model flux is then median smoothened with a kernel of 101 pixels, and fit with a Chebyshev polynomial after masking the Paschen lines, similar to before. This gives a more unbiased estimate of the instrument response than the A0 spectrum, primarily because the Paschen lines in 16 Cyg B are not broadened to the same extent as the A0 stars, with much lesser rotational, temperature, and pressure broadening \citep{gray_observation_1992}.

This procedure gives us the relative instrument response which corrects for the inter, and intra-order response variation, and converts the measured HPF spectrum from electrons to spectral flux density units. To account for varying  atmospheric extinction due to changing water vapour and dust levels between the epoch corresponding to the flare observations, and that for the 16 Cyg B observations, we conservatively assume a $\sim 10\%$ error. This is based on observations of CALSPEC flux standards to estimate the HPF instrument response, where we note relative chromatic trends at $\sim 10\%$ across the response calculated for the HPF bandpass across observations that are temporally spaced by more than a few minutes. Additionally, we also include a  10\% error term to account for inaccuracies in the inter-order normalization while correcting the instrument response. Adding these individual error terms in quadrature, we ascribe an error of $15\%$ to our response corrections per extracted pixel.

\subsection{Absolute calibration}\label{sec:absolute}
We also scale the response corrected vB 10 spectrum to $3.435 \times 10^{-14}$ \specificflux{} corresponding to its 2MASS J magnitude of 9.9  at 12350 \AA~\citep{cutri_2mass_2003}. Doing this enables us to estimate the absolute flux (and therefore energy) emitted in specific atomic lines during the flare. Here we make the assumption that the bolometric luminosity of vB 10 remains constant between the epoch of 2MASS observations in 2003 \citep{cutri_2mass_2003}, and the vB 10 flares as observed by HPF. 

\subsection{Spectral subtraction}\label{sec:spectralsubtraction}
The complexity of the late M dwarf spectrum, and lack of real continuum for any equivalent-width like measurements, led us to adopt the spectral subtraction method to quantify the emission flux during the flare \citep{paulson_optical_2006}. We normalize and take the weighted average for 2x 945 s exposures of vB 10 from 2018 September 24, 03:19 UTC in the stellar rest frame, as a quiescent epoch reference spectrum. This spectrum is telluric corrected similar to the flare spectrum described in Section \ref{sec:hpf}. \replaced{In addition}{Before subtraction}, both the flare and quiescent spectrum are normalized, instrument response corrected and flux calibrated using the procedure described in Section \ref{sec:relative} and Section \ref{sec:absolute}. \added{An example of this is shown for the Pa 7 ($\delta$) line in Figure \ref{fig:SpectralSubtraction}, where we note that the Pa (and similarly the Ca or He) lines can get obscured in the molecular haze for such an ultracool dwarf, and are not observed to be in strong absorption during quiescence.}. Since HPF is a highly stabilized spectrograph with a scrambled fiber input \citep{halverson_efficient_2015}, we can perform this subtraction without accruing errors from variable seeing conditions (between the reference and flare epochs) or guiding jitter \citep{kanodia_harsh_2021}.

\section{ACAM photometry reduction}\label{appsec:acam}

All ACAM frames were 2D bias and dark subtracted. In addition, we reject cosmic rays using the L.A. Cosmic technique \citep{van_dokkum_cosmic-ray_2001} which uses the Laplacian edge detection as implemented in \texttt{ccdproc} \citep{craig_astropyccdproc_2017}. We also correct for column pattern noise in the detector, by masking out the sources, and then subtracting the masked mean value from each column.  An approximate world coordinate system (WCS) is available in the header in the form of the telescope right ascension (RA) and declination, but that does not take into account the orientation of the guide camera. In the case of the ACAM, we find that packages like \texttt{astrometry.net} \citep{lang_astrometrynet_2010} fail in finding a plate solution due to a combination of the small field of view, limiting magnitude, and the movement of high proper motion targets in the field. A more accurate WCS plate solution is added to the headers by detecting sources in the images, and comparing them to the \gaia{} DR2 catalogue \citep{gaia_collaboration_gaia_2018} within a 5' radius using \texttt{astroquery} \citep{ginsburg_astroquery_2019}. This subset of the \gaia{} catalogue is propagated to the epoch of observation by applying space motion, and a transformation from the detected pixel positions to \gaia{} positions is applied to obtain a WCS solution. These images were filtered out by eye based on threshold count values described below. In addition, we also filtered out double images with  a moving FOV,  low peak counts in the target star ($<$ 7000 counts), and background sky over exposure ($>$ 1000 counts). A multi-aperture photometric data analysis was completed using the Java-based image processing software AstroImageJ \citep[AIJ;][]{collins_astroimagej_2017}. We use a photometric aperture of 12 pixels (3.25$\arcsec$) with an inner sky annulus of 21 pixels (5.69$\arcsec$) and outer sky annulus of 32 pixels (8.67$\arcsec$) were selected based on the PSF FWHM of the target star.

\section{Acknowledgements}

We thank the anonymous referees for their helpful feedback. 

This research has made use of the SIMBAD database, operated at CDS, Strasbourg, France, 
and NASA's Astrophysics Data System Bibliographic Services.

Computations for this research were performed on the Pennsylvania State University’s Institute for Computational and Data Sciences Advanced CyberInfrastructure (ICDS-ACI), including the CyberLAMP cluster supported by NSF grant MRI-1626251.

The Center for Exoplanets and Habitable Worlds is supported by the Pennsylvania State University, the Eberly College of Science, and the Pennsylvania Space Grant Consortium. 

This work has made use of data from the European Space Agency (ESA) mission
{\it Gaia} (\url{https://www.cosmos.esa.int/gaia}), processed by the {\it Gaia}
Data Processing and Analysis Consortium (DPAC,
\url{https://www.cosmos.esa.int/web/gaia/dpac/consortium}). Funding for the DPAC
has been provided by national institutions, in particular the institutions
participating in the {\it Gaia} Multilateral Agreement.

These results are based on observations obtained with the Habitable-zone Planet Finder Spectrograph on the HET. We acknowledge support from NSF grants
AST-1006676, AST-1126413, AST-1310885, AST-1310875, AST-1910954, AST-1907622, AST-1909506, ATI 2009889, ATI-2009982, AST-2108512, and the NASA Astrobiology Institute (NNA09DA76A) in the pursuit of precision radial velocities in the NIR. The HPF team also acknowledges support from the Heising-Simons Foundation via grant 2017-0494.  The Hobby-Eberly Telescope is a joint project of the University of Texas at Austin, the Pennsylvania State University, Ludwig-Maximilians-Universität München, and Georg-August Universität Gottingen. The HET is named in honor of its principal benefactors, William P. Hobby and Robert E. Eberly. The HET collaboration acknowledges the support and resources from the Texas Advanced Computing Center. We thank the Resident astronomers and Telescope Operators at the HET for the skillful execution of our observations with HPF. We would like to acknowledge that the HET is built on Indigenous land. Moreover, we would like to acknowledge and pay our respects to the Carrizo \& Comecrudo, Coahuiltecan, Caddo, Tonkawa, Comanche, Lipan Apache, Alabama-Coushatta, Kickapoo, Tigua Pueblo, and all the American Indian and Indigenous Peoples and communities who have been or have become a part of these lands and territories in Texas, here on Turtle Island.

This work includes data from 2MASS, which is a joint project of the University of Massachusetts and IPAC at Caltech funded by NASA and the NSF.  
CIC acknowledges support by NASA Headquarters under the NASA Earth and Space Science Fellowship Program through grant 80NSSC18K1114.  
SK would like to acknowledge Annie Clark and Theodora for help with this project.

\facilities{\gaia{}, HET (HPF)}
\software{
AstroImageJ \citep{collins_astroimagej_2017}, 
\texttt{astropy} \citep{robitaille_astropy_2013, astropy_collaboration_astropy_2018},
\texttt{astroquery} \citep{ginsburg_astroquery_2019}, 
\texttt{barycorrpy} \citep{kanodia_python_2018}, 
\texttt{ccdproc} \citep{craig_astropyccdproc_2017}, 
\texttt{HxRGproc} \citep{ninan_habitable-zone_2018},
\texttt{ipython} \citep{perez_ipython_2007},
\texttt{matplotlib} \citep{hunter_matplotlib_2007},
\texttt{numpy} \citep{oliphant_numpy_2006},
\texttt{pandas} \citep{mckinney_data_2010},
\texttt{photutils} \citep{bradley_astropyphotutils_2020}
\texttt{scipy} \citep{oliphant_python_2007, virtanen_scipy_2020},
\texttt{telfit} \citep{gullikson_correcting_2014}
\texttt{terraspec} \citep{bender_sdss-het_2012, lockwood_near-ir_2014}
}

\bibliography{references}

\begin{thebibliography}{}
\expandafter\ifx\csname natexlab\endcsname\relax\def\natexlab#1{#1}\fi
\providecommand{\url}[1]{\href{#1}{#1}}

\bibitem[{Allard {et~al.}(2012)Allard, Homeier, \&
  Freytag}]{allard_models_2012}
Allard, F., Homeier, D., \& Freytag, B. 2012, Philosophical Transactions of the
  Royal Society of London Series A, 370, 2765.
\newblock \url{http://adsabs.harvard.edu/abs/2012RSPTA.370.2765A}

\bibitem[{Antolin \& Rouppe van~der Voort(2012)}]{antolin_observing_2012}
Antolin, P., \& Rouppe van~der Voort, L. 2012, The Astrophysical Journal, 745,
  152.
\newblock \url{https://ui.adsabs.harvard.edu/abs/2012ApJ...745..152A}

\bibitem[{{Astropy Collaboration} {et~al.}(2018){Astropy Collaboration},
  Price-Whelan, Sipőcz, Günther, Lim, Crawford, Conseil, Shupe, Craig,
  Dencheva, Ginsburg, VanderPlas, Bradley, Pérez-Suárez, de~Val-Borro,
  Aldcroft, Cruz, Robitaille, Tollerud, Ardelean, Babej, Bach, Bachetti,
  Bakanov, Bamford, Barentsen, Barmby, Baumbach, Berry, Biscani, Boquien,
  Bostroem, Bouma, Brammer, Bray, Breytenbach, Buddelmeijer, Burke, Calderone,
  Cano~Rodríguez, Cara, Cardoso, Cheedella, Copin, Corrales, Crichton,
  D'Avella, Deil, Depagne, Dietrich, Donath, Droettboom, Earl, Erben, Fabbro,
  Ferreira, Finethy, Fox, Garrison, Gibbons, Goldstein, Gommers, Greco,
  Greenfield, Groener, Grollier, Hagen, Hirst, Homeier, Horton, Hosseinzadeh,
  Hu, Hunkeler, Ivezić, Jain, Jenness, Kanarek, Kendrew, Kern, Kerzendorf,
  Khvalko, King, Kirkby, Kulkarni, Kumar, Lee, Lenz, Littlefair, Ma, Macleod,
  Mastropietro, McCully, Montagnac, Morris, Mueller, Mumford, Muna, Murphy,
  Nelson, Nguyen, Ninan, Nöthe, Ogaz, Oh, Parejko, Parley, Pascual, Patil,
  Patil, Plunkett, Prochaska, Rastogi, Reddy~Janga, Sabater, Sakurikar,
  Seifert, Sherbert, Sherwood-Taylor, Shih, Sick, Silbiger, Singanamalla,
  Singer, Sladen, Sooley, Sornarajah, Streicher, Teuben, Thomas, Tremblay,
  Turner, Terrón, van Kerkwijk, de~la Vega, Watkins, Weaver, Whitmore,
  Woillez, Zabalza, \& {Astropy
  Contributors}}]{astropy_collaboration_astropy_2018}
{Astropy Collaboration}, Price-Whelan, A.~M., Sipőcz, B.~M., {et~al.} 2018,
  The Astronomical Journal, 156, 123.
\newblock \url{https://ui.adsabs.harvard.edu/abs/2018AJ....156..123A}

\bibitem[{Batalha {et~al.}(2019)Batalha, Lewis, Fortney, Batalha, Kempton,
  Lewis, \& Line}]{batalha_precision_2019}
Batalha, N.~E., Lewis, T., Fortney, J.~J., {et~al.} 2019, The Astrophysical
  Journal Letters, 885, L25.
\newblock \url{http://adsabs.harvard.edu/abs/2019ApJ...885L..25B}

\bibitem[{Bender {et~al.}(2012)Bender, Mahadevan, Deshpande, Wright, Roy,
  Terrien, Sigurdsson, Ramsey, Schneider, \& Fleming}]{bender_sdss-het_2012}
Bender, C.~F., Mahadevan, S., Deshpande, R., {et~al.} 2012, The Astrophysical
  Journal, 751, L31.
\newblock \url{https://ui.adsabs.harvard.edu/abs/2012ApJ...751L..31B}

\bibitem[{Berger {et~al.}(2008)Berger, Basri, Gizis, Giampapa, Rutledge,
  Liebert, Martín, Fleming, Johns‐Krull, Phan‐Bao, \&
  Sherry}]{berger_simultaneous_2008-1}
Berger, E., Basri, G., Gizis, J.~E., {et~al.} 2008, The Astrophysical Journal,
  676, 1307.
\newblock \url{http://stacks.iop.org/0004-637X/676/i=2/a=1307}

\bibitem[{Bohlin {et~al.}(2014)Bohlin, Gordon, \&
  Tremblay}]{bohlin_techniques_2014}
Bohlin, R.~C., Gordon, K.~D., \& Tremblay, P.-E. 2014, Publications of the
  Astronomical Society of the Pacific, 126, 711.
\newblock \url{http://adsabs.harvard.edu/abs/2014PASP..126..711B}

\bibitem[{Bohlin {et~al.}(2020)Bohlin, Hubeny, \& Rauch}]{bohlin_new_2020}
Bohlin, R.~C., Hubeny, I., \& Rauch, T. 2020, The Astronomical Journal, 160,
  21.
\newblock \url{http://adsabs.harvard.edu/abs/2020AJ....160...21B}

\bibitem[{Bradley {et~al.}(2020)Bradley, Sipőcz, Robitaille, Tollerud,
  Vinícius, Deil, Barbary, Wilson, Busko, Günther, Cara, Conseil, Bostroem,
  Droettboom, Bray, Bratholm, Lim, Barentsen, Craig, Pascual, Perren, Greco,
  Donath, Val-Borro, Kerzendorf, Bach, Weaver, D'Eugenio, Souchereau, \&
  Ferreira}]{bradley_astropyphotutils_2020}
Bradley, L., Sipőcz, B., Robitaille, T., {et~al.} 2020, astropy/photutils:
  1.0.0,  Zenodo, doi:10.5281/zenodo.4044744.
\newblock \url{https://doi.org/10.5281/zenodo.4044744}

\bibitem[{Canfield {et~al.}(1990)Canfield, Penn, Wulser, \&
  Kiplinger}]{canfield_h-alpha_1990}
Canfield, R.~C., Penn, M.~J., Wulser, J.-P., \& Kiplinger, A.~L. 1990, The
  Astrophysical Journal, 363, 318.
\newblock \url{http://adsabs.harvard.edu/abs/1990ApJ...363..318C}

\bibitem[{Castelli \& Kurucz(1994)}]{castelli_model_1994}
Castelli, F., \& Kurucz, R.~L. 1994, Astronomy and Astrophysics, 281, 817.
\newblock \url{http://adsabs.harvard.edu/abs/1994A%26A...281..817C}

\bibitem[{Chang {et~al.}(2015)Chang, Byun, \& Hartman}]{chang_photometric_2015}
Chang, S.~W., Byun, Y.~I., \& Hartman, J.~D. 2015, The Astrophysical Journal,
  814, 35.
\newblock \url{https://ui.adsabs.harvard.edu/abs/2015ApJ...814...35C}

\bibitem[{Chebyshev(1854)}]{chebyshev_theorie_1854}
Chebyshev, P.~L. 1854, Mémoires Présentés a l'Académie Impériale des
  Sciences de St. Pétersbourg par Divers Savants, 539.
\newblock \url{https://www.math.technion.ac.il/hat/fpapers/cheb11.pdf}

\bibitem[{Clough {et~al.}(2005)Clough, Shephard, Mlawer, Delamere, Iacono,
  Cady-Pereira, Boukabara, \& Brown}]{clough_atmospheric_2005}
Clough, S.~A., Shephard, M.~W., Mlawer, E.~J., {et~al.} 2005, Journal of
  Quantitative Spectroscopy and Radiative Transfer, 91, 233.
\newblock \url{http://adsabs.harvard.edu/abs/2005JQSRT..91..233C}

\bibitem[{Collaboration {et~al.}(2021)Collaboration, Brown, Vallenari, Prusti,
  de~Bruijne, Babusiaux, Biermann, Creevey, Evans, Eyer, Hutton, Jansen, Jordi,
  Klioner, Lammers, Lindegren, Luri, Mignard, Panem, Pourbaix, Randich,
  Sartoretti, Soubiran, Walton, Arenou, Bailer-Jones, Bastian, Cropper,
  Drimmel, Katz, Lattanzi, van Leeuwen, Bakker, Cacciari, Castañeda,
  De~Angeli, Ducourant, Fabricius, Fouesneau, Frémat, Guerra, Guerrier,
  Guiraud, Jean-Antoine~Piccolo, Masana, Messineo, Mowlavi, Nicolas,
  Nienartowicz, Pailler, Panuzzo, Riclet, Roux, Seabroke, Sordo, Tanga,
  Thévenin, Gracia-Abril, Portell, Teyssier, Altmann, Andrae, Bellas-Velidis,
  Benson, Berthier, Blomme, Brugaletta, Burgess, Busso, Carry, Cellino, Cheek,
  Clementini, Damerdji, Davidson, Delchambre, Dell'Oro, Fernández-Hernández,
  Galluccio, García-Lario, Garcia-Reinaldos, González-Núñez, Gosset,
  Haigron, Halbwachs, Hambly, Harrison, Hatzidimitriou, Heiter, Hernández,
  Hestroffer, Hodgkin, Holl, Janßen, Jevardat~de Fombelle, Jordan,
  Krone-Martins, Lanzafame, Löffler, Lorca, Manteiga, Marchal, Marrese,
  Moitinho, Mora, Muinonen, Osborne, Pancino, Pauwels, Petit, Recio-Blanco,
  Richards, Riello, Rimoldini, Robin, Roegiers, Rybizki, Sarro, Siopis, Smith,
  Sozzetti, Ulla, Utrilla, van Leeuwen, van Reeven, Abbas, Abreu~Aramburu,
  Accart, Aerts, Aguado, Ajaj, Altavilla, Álvarez, Álvarez Cid-Fuentes,
  Alves, Anderson, Anglada~Varela, Antoja, Audard, Baines, Baker,
  Balaguer-Núñez, Balbinot, Balog, Barache, Barbato, Barros, Barstow,
  Bartolomé, Bassilana, Bauchet, Baudesson-Stella, Becciani, Bellazzini,
  Bernet, Bertone, Bianchi, Blanco-Cuaresma, Boch, Bombrun, Bossini,
  Bouquillon, Bragaglia, Bramante, Breedt, Bressan, Brouillet, Bucciarelli,
  Burlacu, Busonero, Butkevich, Buzzi, Caffau, Cancelliere, Cánovas,
  Cantat-Gaudin, Carballo, Carlucci, Carnerero, Carrasco, Casamiquela,
  Castellani, Castro-Ginard, Castro~Sampol, Chaoul, Charlot, Chemin, Chiavassa,
  Cioni, Comoretto, Cooper, Cornez, Cowell, Crifo, Crosta, Crowley, Dafonte,
  Dapergolas, David, David, de~Laverny, De~Luise, De~March, De~Ridder,
  de~Souza, de~Teodoro, de~Torres, del Peloso, del Pozo, Delbo, Delgado,
  Delgado, Delisle, Di~Matteo, Diakite, Diener, Distefano, Dolding, Eappachen,
  Edvardsson, Enke, Esquej, Fabre, Fabrizio, Faigler, Fedorets, Fernique,
  Fienga, Figueras, Fouron, Fragkoudi, Fraile, Franke, Gai, Garabato,
  Garcia-Gutierrez, García-Torres, Garofalo, Gavras, Gerlach, Geyer, Giacobbe,
  Gilmore, Girona, Giuffrida, Gomel, Gomez, Gonzalez-Santamaria,
  González-Vidal, Granvik, Gutiérrez-Sánchez, Guy, Hauser, Haywood, Helmi,
  Hidalgo, Hilger, Hładczuk, Hobbs, Holland, Huckle, Jasniewicz, Jonker,
  Juaristi~Campillo, Julbe, Karbevska, Kervella, Khanna, Kochoska, Kontizas,
  Kordopatis, Korn, Kostrzewa-Rutkowska, Kruszyńska, Lambert, Lanza, Lasne,
  Le~Campion, Le~Fustec, Lebreton, Lebzelter, Leccia, Leclerc, Lecoeur-Taibi,
  Liao, Licata, Lindstrøm, Lister, Livanou, Lobel, Madrero~Pardo, Managau,
  Mann, Marchant, Marconi, Marcos~Santos, Marinoni, Marocco, Marshall,
  Martin~Polo, Martín-Fleitas, Masip, Massari, Mastrobuono-Battisti, Mazeh,
  McMillan, Messina, Michalik, Millar, Mints, Molina, Molinaro, Molnár,
  Montegriffo, Mor, Morbidelli, Morel, Morris, Mulone, Munoz, Muraveva, Murphy,
  Musella, Noval, Ordénovic, Orrù, Osinde, Pagani, Pagano, Palaversa,
  Palicio, Panahi, Pawlak, Peñalosa~Esteller, Penttilä, Piersimoni, Pineau,
  Plachy, Plum, Poggio, Poretti, Poujoulet, Prša, Pulone, Racero, Ragaini,
  Rainer, Raiteri, Rambaux, Ramos, Ramos-Lerate, Re~Fiorentin, Regibo, Reylé,
  Ripepi, Riva, Rixon, Robichon, Robin, Roelens, Rohrbasser, Romero-Gómez,
  Rowell, Royer, Rybicki, Sadowski, Sagristà~Sellés, Sahlmann, Salgado,
  Salguero, Samaras, Sanchez~Gimenez, Sanna, Santoveña, Sarasso, Schultheis,
  Sciacca, Segol, Segovia, Ségransan, Semeux, Shahaf, Siddiqui, Siebert,
  Siltala, Slezak, Smart, Solano, Solitro, Souami, Souchay, Spagna, Spoto,
  Steele, Steidelmüller, Stephenson, Süveges, Szabados, Szegedi-Elek, Taris,
  Tauran, Taylor, Teixeira, Thuillot, Tonello, Torra, Torra, Turon, Unger,
  Vaillant, van Dillen, Vanel, Vecchiato, Viala, Vicente, Voutsinas, Weiler,
  Wevers, Wyrzykowski, Yoldas, Yvard, Zhao, Zorec, Zucker, Zurbach, \&
  Zwitter}]{collaboration_gaia_2021}
Collaboration, G., Brown, A. G.~A., Vallenari, A., {et~al.} 2021, Astronomy and
  Astrophysics, 649, A1.
\newblock \url{https://ui.adsabs.harvard.edu/abs/2021A&A...649A...1G/abstract}

\bibitem[{Collins {et~al.}(2017)Collins, Kielkopf, Stassun, \&
  Hessman}]{collins_astroimagej_2017}
Collins, K.~A., Kielkopf, J.~F., Stassun, K.~G., \& Hessman, F.~V. 2017, The
  Astronomical Journal, 153, 77.
\newblock \url{http://adsabs.harvard.edu/abs/2017AJ....153...77C}

\bibitem[{Craig {et~al.}(2017)Craig, Crawford, Seifert, Robitaille, Sipőcz,
  Walawender, Vinícius, Ninan, Droettboom, Youn, Tollerud, Bray, Walker,
  Janga, Stotts, Günther, Rol, Bach, Bradley, Deil, Price-Whelan, Barbary,
  Horton, Schoenell, Heidt, Gasdia, Nelson, \&
  Streicher}]{craig_astropyccdproc_2017}
Craig, M., Crawford, S., Seifert, M., {et~al.} 2017, astropy/ccdproc:
  v1.3.0.post1, doi:10.5281/zenodo.1069648.
\newblock \url{https://doi.org/10.5281/zenodo.1069648}

\bibitem[{Crespo-Chacón {et~al.}(2006)Crespo-Chacón, Montes, García-Alvarez,
  Fernández-Figueroa, López-Santiago, \& Foing}]{crespo-chacon_analysis_2006}
Crespo-Chacón, I., Montes, D., García-Alvarez, D., {et~al.} 2006, Astronomy
  and Astrophysics, Volume 452, Issue 3, June IV 2006, pp.987-1000, 452, 987.
\newblock
  \url{https://ui.adsabs.harvard.edu/abs/2006A%26A...452..987C/abstract}

\bibitem[{Cuntz \& Guinan(2016)}]{cuntz_about_2016}
Cuntz, M., \& Guinan, E.~F. 2016, The Astrophysical Journal, 827, 79.
\newblock \url{https://ui.adsabs.harvard.edu/abs/2016ApJ...827...79C}

\bibitem[{Cutri {et~al.}(2003)Cutri, Skrutskie, van Dyk, Beichman, Carpenter,
  Chester, Cambresy, Evans, Fowler, Gizis, Howard, Huchra, Jarrett, Kopan,
  Kirkpatrick, Light, Marsh, McCallon, Schneider, Stiening, Sykes, Weinberg,
  Wheaton, Wheelock, \& Zacarias}]{cutri_2mass_2003}
Cutri, R.~M., Skrutskie, M.~F., van Dyk, S., {et~al.} 2003, "The IRSA 2MASS
  All-Sky Point Source Catalog, NASA/IPAC Infrared Science Archive.
  http://irsa.ipac.caltech.edu/applications/Gator/".
\newblock \url{http://adsabs.harvard.edu/abs/2003tmc..book.....C}

\bibitem[{Donati {et~al.}(2020)Donati, Kouach, Moutou, Doyon, Delfosse,
  Artigau, Baratchart, Lacombe, Barrick, Hébrard, Bouchy, Saddlemyer, Parès,
  Rabou, Micheau, Dolon, Reshetov, Challita, Carmona, Striebig, Thibault,
  Martioli, Cook, Fouqué, Vermeulen, Wang, Arnold, Pepe, Boisse, Figueira,
  Bouvier, Ray, Feugeade, Morin, Alencar, Hobson, Castilho, Udry, Santos,
  Hernandez, Benedict, Vallée, Gallou, Dupieux, Larrieu, Perruchot, Sottile,
  Moreau, Usher, Baril, Wildi, Chazelas, Malo, Bonfils, Loop, Kerley, Wevers,
  Dunn, Pazder, Macdonald, Dubois, Carrié, Valentin, Henault, Yan, \&
  Steinmetz}]{donati_spirou_2020}
Donati, J.~F., Kouach, D., Moutou, C., {et~al.} 2020, Monthly Notices of the
  Royal Astronomical Society, 498, 5684, aDS Bibcode: 2020MNRAS.498.5684D.
\newblock \url{https://ui.adsabs.harvard.edu/abs/2020MNRAS.498.5684D}

\bibitem[{Dressing \& Charbonneau(2015)}]{dressing_occurrence_2015}
Dressing, C.~D., \& Charbonneau, D. 2015, The Astrophysical Journal, 807, 45.
\newblock \url{http://adsabs.harvard.edu/abs/2015ApJ...807...45D}

\bibitem[{Díaz {et~al.}(2011)Díaz, González, Levato, \&
  Grosso}]{diaz_accurate_2011}
Díaz, C.~G., González, J.~F., Levato, H., \& Grosso, M. 2011, Astronomy and
  Astrophysics, 531, A143.
\newblock \url{http://adsabs.harvard.edu/abs/2011A%26A...531A.143D}

\bibitem[{Endl {et~al.}(2000)Endl, Kürster, \& Els}]{endl_planet_2000}
Endl, M., Kürster, M., \& Els, S. 2000, Astronomy and Astrophysics, v.362,
  p.585-594 (2000), 362, 585.
\newblock
  \url{https://ui.adsabs.harvard.edu/abs/2000A%26A...362..585E/abstract}

\bibitem[{Erkaev {et~al.}(2007)Erkaev, Kulikov, Lammer, Selsis, Langmayr,
  Jaritz, \& Biernat}]{erkaev_roche_2007}
Erkaev, N.~V., Kulikov, Y.~N., Lammer, H., {et~al.} 2007, Astronomy and
  Astrophysics, 472, 329.
\newblock \url{https://ui.adsabs.harvard.edu/abs/2007A&A...472..329E/abstract}

\bibitem[{Feinstein {et~al.}(2020)Feinstein, Montet, Ansdell, Nord, Bean,
  Günther, Gully-Santiago, \& Schlieder}]{feinstein_flare_2020}
Feinstein, A.~D., Montet, B.~T., Ansdell, M., {et~al.} 2020, The Astronomical
  Journal, 160, 219.
\newblock \url{https://ui.adsabs.harvard.edu/abs/2020AJ....160..219F}

\bibitem[{Fleming {et~al.}(2003)Fleming, Giampapa, \&
  Garza}]{fleming_quiescent_2003}
Fleming, T.~A., Giampapa, M.~S., \& Garza, D. 2003, The Astrophysical Journal,
  594, 982.
\newblock \url{http://adsabs.harvard.edu/abs/2003ApJ...594..982F}

\bibitem[{Fleming {et~al.}(2000)Fleming, Giampapa, \&
  Schmitt}]{fleming_x-ray_2000}
Fleming, T.~A., Giampapa, M.~S., \& Schmitt, J. H. M.~M. 2000, The
  Astrophysical Journal, 533, 372.
\newblock \url{http://adsabs.harvard.edu/abs/2000ApJ...533..372F}

\bibitem[{Fouqué {et~al.}(2018)Fouqué, Moutou, Malo, Martioli, Lim,
  Rajpurohit, Artigau, Delfosse, Donati, Forveille, Morin, Allard, Delage,
  Doyon, Hébrard, \& Neves}]{fouque_spirou_2018}
Fouqué, P., Moutou, C., Malo, L., {et~al.} 2018, Monthly Notices of the Royal
  Astronomical Society, 475, 1960.
\newblock \url{https://ui.adsabs.harvard.edu/abs/2018MNRAS.475.1960F}

\bibitem[{Fuhrmeister {et~al.}(2011)Fuhrmeister, Lalitha, Poppenhaeger, Rudolf,
  Liefke, Reiners, Schmitt, \& Ness}]{fuhrmeister_multi-wavelength_2011}
Fuhrmeister, B., Lalitha, S., Poppenhaeger, K., {et~al.} 2011, Astronomy \&amp;
  Astrophysics, Volume 534, id.A133,
  {\textless}NUMPAGES{\textgreater}17{\textless}/NUMPAGES{\textgreater} pp.,
  534, A133.
\newblock
  \url{https://ui.adsabs.harvard.edu/abs/2011A%26A...534A.133F/abstract}

\bibitem[{Fuhrmeister {et~al.}(2007)Fuhrmeister, Liefke, \&
  Schmitt}]{fuhrmeister_simultaneous_2007}
Fuhrmeister, B., Liefke, C., \& Schmitt, J. H. M.~M. 2007, Astronomy and
  Astrophysics, Volume 468, Issue 1, June II 2007, pp.221-231, 468, 221.
\newblock
  \url{https://ui.adsabs.harvard.edu/abs/2007A%26A...468..221F/abstract}

\bibitem[{Fuhrmeister {et~al.}(2008)Fuhrmeister, Liefke, Schmitt, \&
  Reiners}]{fuhrmeister_multiwavelength_2008}
Fuhrmeister, B., Liefke, C., Schmitt, J. H. M.~M., \& Reiners, A. 2008,
  Astronomy and Astrophysics, 487, 293.
\newblock \url{http://adsabs.harvard.edu/abs/2008A%26A...487..293F}

\bibitem[{Fuhrmeister {et~al.}(2018)Fuhrmeister, Czesla, Schmitt, Jeffers,
  Caballero, Zechmeister, Reiners, Ribas, Amado, Quirrenbach, Béjar,
  Galadí-Enríquez, Guenther, Kürster, Montes, \&
  Seifert}]{fuhrmeister_carmenes_2018}
Fuhrmeister, B., Czesla, S., Schmitt, J. H. M.~M., {et~al.} 2018, Astronomy \&
  Astrophysics, 615, A14, publisher: EDP Sciences.
\newblock
  \url{https://www.aanda.org/articles/aa/abs/2018/07/aa32204-17/aa32204-17.html}

\bibitem[{Fuhrmeister {et~al.}(2019)Fuhrmeister, Czesla, Schmitt, Johnson,
  Schöfer, Jeffers, Caballero, Zechmeister, Reiners, Ribas, Amado,
  Quirrenbach, Bauer, Béjar, Cortés-Contreras, Díez~Alonso, Dreizler,
  Galadí-Enríquez, Guenther, Kaminski, Kürster, Lafarga, \&
  Montes}]{fuhrmeister_carmenes_2019}
---. 2019, Astronomy and Astrophysics, 623, A24.
\newblock \url{https://ui.adsabs.harvard.edu/abs/2019A&A...623A..24F/abstract}

\bibitem[{Fuhrmeister {et~al.}(2020)Fuhrmeister, Czesla, Hildebrandt, Nagel,
  Schmitt, Jeffers, Caballero, Hintz, Johnson, Schöfer, Zechmeister, Reiners,
  Ribas, Amado, Quirrenbach, Nortmann, Bauer, Béjar, Cortés-Contreras,
  Dreizler, Galadí-Enríquez, Hatzes, Kaminski, Kürster, Lafarga, \&
  Montes}]{fuhrmeister_carmenes_2020}
Fuhrmeister, B., Czesla, S., Hildebrandt, L., {et~al.} 2020, arXiv:2006.09372
  [astro-ph], arXiv: 2006.09372.
\newblock \url{http://arxiv.org/abs/2006.09372}

\bibitem[{Fulton {et~al.}(2017)Fulton, Petigura, Howard, Isaacson, Marcy,
  Cargile, Hebb, Weiss, Johnson, Morton, Sinukoff, Crossfield, \&
  Hirsch}]{fulton_california-kepler_2017}
Fulton, B.~J., Petigura, E.~A., Howard, A.~W., {et~al.} 2017, The Astronomical
  Journal, 154, 109.
\newblock \url{https://ui.adsabs.harvard.edu/abs/2017AJ....154..109F/abstract}

\bibitem[{{Gaia Collaboration} {et~al.}(2018){Gaia Collaboration}, Brown,
  Vallenari, Prusti, de~Bruijne, Babusiaux, Bailer-Jones, Biermann, Evans,
  Eyer, Jansen, Jordi, Klioner, Lammers, Lindegren, Luri, Mignard, Panem,
  Pourbaix, Randich, Sartoretti, Siddiqui, Soubiran, van Leeuwen, Walton,
  Arenou, Bastian, Cropper, Drimmel, Katz, Lattanzi, Bakker, Cacciari,
  Castañeda, Chaoul, Cheek, De~Angeli, Fabricius, Guerra, Holl, Masana,
  Messineo, Mowlavi, Nienartowicz, Panuzzo, Portell, Riello, Seabroke, Tanga,
  Thévenin, Gracia-Abril, Comoretto, Garcia-Reinaldos, Teyssier, Altmann,
  Andrae, Audard, Bellas-Velidis, Benson, Berthier, Blomme, Burgess, Busso,
  Carry, Cellino, Clementini, Clotet, Creevey, Davidson, De~Ridder, Delchambre,
  Dell'Oro, Ducourant, Fernández-Hernández, Fouesneau, Frémat, Galluccio,
  García-Torres, González-Núñez, González-Vidal, Gosset, Guy, Halbwachs,
  Hambly, Harrison, Hernández, Hestroffer, Hodgkin, Hutton, Jasniewicz,
  Jean-Antoine-Piccolo, Jordan, Korn, Krone-Martins, Lanzafame, Lebzelter,
  Löffler, Manteiga, Marrese, Martín-Fleitas, Moitinho, Mora, Muinonen,
  Osinde, Pancino, Pauwels, Petit, Recio-Blanco, Richards, Rimoldini, Robin,
  Sarro, Siopis, Smith, Sozzetti, Süveges, Torra, van Reeven, Abbas,
  Abreu~Aramburu, Accart, Aerts, Altavilla, Álvarez, Alvarez, Alves, Anderson,
  Andrei, Anglada~Varela, Antiche, Antoja, Arcay, Astraatmadja, Bach, Baker,
  Balaguer-Núñez, Balm, Barache, Barata, Barbato, Barblan, Barklem, Barrado,
  Barros, Barstow, Bartholomé~Muñoz, Bassilana, Becciani, Bellazzini,
  Berihuete, Bertone, Bianchi, Bienaymé, Blanco-Cuaresma, Boch, Boeche,
  Bombrun, Borrachero, Bossini, Bouquillon, Bourda, Bragaglia, Bramante,
  Breddels, Bressan, Brouillet, Brüsemeister, Brugaletta, Bucciarelli,
  Burlacu, Busonero, Butkevich, Buzzi, Caffau, Cancelliere, Cannizzaro,
  Cantat-Gaudin, Carballo, Carlucci, Carrasco, Casamiquela, Castellani,
  Castro-Ginard, Charlot, Chemin, Chiavassa, Cocozza, Costigan, Cowell, Crifo,
  Crosta, Crowley, Cuypers, Dafonte, Damerdji, Dapergolas, David, David,
  de~Laverny, De~Luise, De~March, de~Martino, de~Souza, de~Torres, Debosscher,
  del Pozo, Delbo, Delgado, Delgado, Di~Matteo, Diakite, Diener, Distefano,
  Dolding, Drazinos, Durán, Edvardsson, Enke, Eriksson, Esquej,
  Eynard~Bontemps, Fabre, Fabrizio, Faigler, Falcão, Farràs~Casas, Federici,
  Fedorets, Fernique, Figueras, Filippi, Findeisen, Fonti, Fraile, Fraser,
  Frézouls, Gai, Galleti, Garabato, García-Sedano, Garofalo, Garralda, Gavel,
  Gavras, Gerssen, Geyer, Giacobbe, Gilmore, Girona, Giuffrida, Glass, Gomes,
  Granvik, Gueguen, Guerrier, Guiraud, Gutiérrez-Sánchez, Haigron,
  Hatzidimitriou, Hauser, Haywood, Heiter, Helmi, Heu, Hilger, Hobbs, Hofmann,
  Holland, Huckle, Hypki, Icardi, Janßen, Jevardat~de Fombelle, Jonker,
  Juhász, Julbe, Karampelas, Kewley, Klar, Kochoska, Kohley, Kolenberg,
  Kontizas, Kontizas, Koposov, Kordopatis, Kostrzewa-Rutkowska, Koubsky,
  Lambert, Lanza, Lasne, Lavigne, Le~Fustec, Le~Poncin-Lafitte, Lebreton,
  Leccia, Leclerc, Lecoeur-Taibi, Lenhardt, Leroux, Liao, Licata, Lindstrøm,
  Lister, Livanou, Lobel, López, Managau, Mann, Mantelet, Marchal, Marchant,
  Marconi, Marinoni, Marschalkó, Marshall, Martino, Marton, Mary, Massari,
  Matijevič, Mazeh, McMillan, Messina, Michalik, Millar, Molina, Molinaro,
  Molnár, Montegriffo, Mor, Morbidelli, Morel, Morris, Mulone, Muraveva,
  Musella, Nelemans, Nicastro, Noval, O'Mullane, Ordénovic, Ordóñez-Blanco,
  Osborne, Pagani, Pagano, Pailler, Palacin, Palaversa, Panahi, Pawlak,
  Piersimoni, Pineau, Plachy, Plum, Poggio, Poujoulet, Prša, Pulone, Racero,
  Ragaini, Rambaux, Ramos-Lerate, Regibo, Reylé, Riclet, Ripepi, Riva, Rivard,
  Rixon, Roegiers, Roelens, Romero-Gómez, Rowell, Royer, Ruiz-Dern, Sadowski,
  Sagristà~Sellés, Sahlmann, Salgado, Salguero, Sanna, Santana-Ros, Sarasso,
  Savietto, Schultheis, Sciacca, Segol, Segovia, Ségransan, Shih, Siltala,
  Silva, Smart, Smith, Solano, Solitro, Sordo, Soria~Nieto, Souchay, Spagna,
  Spoto, Stampa, Steele, Steidelmüller, Stephenson, Stoev, Suess, Surdej,
  Szabados, Szegedi-Elek, Tapiador, Taris, Tauran, Taylor, Teixeira, Terrett,
  Teyssandier, Thuillot, Titarenko, Torra~Clotet, Turon, Ulla, Utrilla, Uzzi,
  Vaillant, Valentini, Valette, van Elteren, Van~Hemelryck, van Leeuwen,
  Vaschetto, Vecchiato, Veljanoski, Viala, Vicente, Vogt, von Essen, Voss,
  Votruba, Voutsinas, Walmsley, Weiler, Wertz, Wevers, Wyrzykowski, Yoldas,
  Žerjal, Ziaeepour, Zorec, Zschocke, Zucker, Zurbach, \&
  Zwitter}]{gaia_collaboration_gaia_2018}
{Gaia Collaboration}, Brown, A. G.~A., Vallenari, A., {et~al.} 2018, Astronomy
  and Astrophysics, 616, A1.
\newblock \url{http://adsabs.harvard.edu/abs/2018A%26A...616A...1G}

\bibitem[{Ginsburg {et~al.}(2019)Ginsburg, Sipőcz, Brasseur, Cowperthwaite,
  Craig, Deil, Guillochon, Guzman, Liedtke, Lian~Lim, Lockhart, Mommert,
  Morris, Norman, Parikh, Persson, Robitaille, Segovia, Singer, Tollerud,
  de~Val-Borro, Valtchanov, Woillez, {Astroquery Collaboration}, \& {a subset
  of astropy Collaboration}}]{ginsburg_astroquery_2019}
Ginsburg, A., Sipőcz, B.~M., Brasseur, C.~E., {et~al.} 2019, The Astronomical
  Journal, 157, 98.
\newblock \url{http://adsabs.harvard.edu/abs/2019AJ....157...98G}

\bibitem[{Gizis {et~al.}(2013)Gizis, Burgasser, Berger, Williams, Vrba, Cruz,
  \& Metchev}]{gizis_kepler_2013}
Gizis, J.~E., Burgasser, A.~J., Berger, E., {et~al.} 2013, The Astrophysical
  Journal, 779, 172, aDS Bibcode: 2013ApJ...779..172G.
\newblock \url{https://ui.adsabs.harvard.edu/abs/2013ApJ...779..172G}

\bibitem[{Gizis {et~al.}(2017)Gizis, Paudel, Mullan, Schmidt, Burgasser, \&
  Williams}]{gizis_k2_2017}
Gizis, J.~E., Paudel, R.~R., Mullan, D., {et~al.} 2017, The Astrophysical
  Journal, 845, 33.
\newblock \url{https://ui.adsabs.harvard.edu/abs/2017ApJ...845...33G}

\bibitem[{Graham \& Cauzzi(2015)}]{graham_temporal_2015}
Graham, D.~R., \& Cauzzi, G. 2015, The Astrophysical Journal, 807, L22, aDS
  Bibcode: 2015ApJ...807L..22G.
\newblock \url{https://ui.adsabs.harvard.edu/abs/2015ApJ...807L..22G}

\bibitem[{Graham {et~al.}(2020)Graham, Cauzzi, Zangrilli, Kowalski, Simões, \&
  Allred}]{graham_spectral_2020}
Graham, D.~R., Cauzzi, G., Zangrilli, L., {et~al.} 2020, The Astrophysical
  Journal, 895, 6.
\newblock \url{http://adsabs.harvard.edu/abs/2020ApJ...895....6G}

\bibitem[{Gray(1992)}]{gray_observation_1992}
Gray, D.~F. 1992, Camb. Astrophys. Ser., Vol. 20,.
\newblock \url{http://adsabs.harvard.edu/abs/1992oasp.book.....G}

\bibitem[{Guedel \& Benz(1993)}]{guedel_x-raymicrowave_1993}
Guedel, M., \& Benz, A.~O. 1993, The Astrophysical Journal, 405, L63, aDS
  Bibcode: 1993ApJ...405L..63G.
\newblock \url{https://ui.adsabs.harvard.edu/abs/1993ApJ...405L..63G}

\bibitem[{Gullikson {et~al.}(2014)Gullikson, Dodson-Robinson, \&
  Kraus}]{gullikson_correcting_2014}
Gullikson, K., Dodson-Robinson, S., \& Kraus, A. 2014, The Astronomical
  Journal, 148, 53.
\newblock \url{http://adsabs.harvard.edu/abs/2014AJ....148...53G}

\bibitem[{Hall \& Ramsey(1992)}]{hall_eclipse_1992}
Hall, J.~C., \& Ramsey, L.~W. 1992, The Astronomical Journal, 104, 1942.
\newblock \url{http://adsabs.harvard.edu/abs/1992AJ....104.1942H}

\bibitem[{Halverson {et~al.}(2015)Halverson, Roy, Mahadevan, Ramsey, Levi,
  Schwab, {Fred Hearty}, \& MacDonald}]{halverson_efficient_2015}
Halverson, S., Roy, A., Mahadevan, S., {et~al.} 2015, The Astrophysical
  Journal, 806, 61.
\newblock \url{http://stacks.iop.org/0004-637X/806/i=1/a=61}

\bibitem[{Hardegree-Ullman {et~al.}(2019)Hardegree-Ullman, Cushing, Muirhead,
  \& Christiansen}]{hardegree-ullman_kepler_2019}
Hardegree-Ullman, K.~K., Cushing, M.~C., Muirhead, P.~S., \& Christiansen,
  J.~L. 2019, The Astronomical Journal, 158, 75.
\newblock \url{http://adsabs.harvard.edu/abs/2019AJ....158...75H}

\bibitem[{Henry {et~al.}(2006)Henry, Jao, Subasavage, Beaulieu, Ianna, Costa,
  \& Méndez}]{henry_solar_2006}
Henry, T.~J., Jao, W.-C., Subasavage, J.~P., {et~al.} 2006, The Astronomical
  Journal, 132, 2360.
\newblock \url{http://adsabs.harvard.edu/abs/2006AJ....132.2360H}

\bibitem[{Herbig(1956)}]{herbig_observations_1956}
Herbig, G.~H. 1956, Publications of the Astronomical Society of the Pacific,
  68, 531.
\newblock \url{http://adsabs.harvard.edu/abs/1956PASP...68..531H}

\bibitem[{Hilton(2011)}]{hilton_galactic_2011}
Hilton, E.~J. 2011, PhD thesis, publication Title: Ph.D. Thesis.
\newblock \url{https://ui.adsabs.harvard.edu/abs/2011PhDT.......144H}

\bibitem[{Honda {et~al.}(2018)Honda, Notsu, Namekata, Notsu, Maehara, Ikuta,
  Nogami, \& Shibata}]{honda_time_2018}
Honda, S., Notsu, Y., Namekata, K., {et~al.} 2018, Publications of the
  Astronomical Society of Japan, 70, doi:10.1093/pasj/psy055, arXiv:
  1804.03771.
\newblock \url{http://arxiv.org/abs/1804.03771}

\bibitem[{Hunter(2007)}]{hunter_matplotlib_2007}
Hunter, J.~D. 2007, Computing in Science Engineering, 9, 90

\bibitem[{Ichimoto \& Kurokawa(1984)}]{ichimoto_h-alpha_1984}
Ichimoto, K., \& Kurokawa, H. 1984, Solar Physics, 93, 105.
\newblock \url{http://adsabs.harvard.edu/abs/1984SoPh...93..105I}

\bibitem[{Irwin {et~al.}(2015)Irwin, Berta-Thompson, Charbonneau, Dittmann,
  Falco, Newton, \& Nutzman}]{irwin_mearth-north_2015}
Irwin, J.~M., Berta-Thompson, Z.~K., Charbonneau, D., {et~al.} 2015, 18, 767,
  conference Name: 18th Cambridge Workshop on Cool Stars, Stellar Systems, and
  the Sun Place: eprint: arXiv:1409.0891.
\newblock \url{http://adsabs.harvard.edu/abs/2015csss...18..767I}

\bibitem[{Johns-Krull {et~al.}(1997)Johns-Krull, Hawley, Basri, \&
  Valenti}]{johns-krull_hamilton_1997}
Johns-Krull, C.~M., Hawley, S.~L., Basri, G., \& Valenti, J.~A. 1997, The
  Astrophysical Journal Supplement Series, 112, 221.
\newblock \url{http://adsabs.harvard.edu/abs/1997ApJS..112..221J}

\bibitem[{Joy \& Humason(1949)}]{joy_observations_1949}
Joy, A.~H., \& Humason, M.~L. 1949, Publications of the Astronomical Society of
  the Pacific, 61, 133.
\newblock \url{https://ui.adsabs.harvard.edu/abs/1949PASP...61..133J}

\bibitem[{Judge {et~al.}(2015)Judge, Kleint, \&
  Sainz~Dalda}]{judge_helium_2015}
Judge, P.~G., Kleint, L., \& Sainz~Dalda, A. 2015, The Astrophysical Journal,
  814, 100, aDS Bibcode: 2015ApJ...814..100J.
\newblock \url{https://ui.adsabs.harvard.edu/abs/2015ApJ...814..100J}

\bibitem[{Kahler {et~al.}(1982)Kahler, Golub, Harnden, Liller, Seward, Vaiana,
  Lovell, Davis, Spencer, Whitehouse, Feldman, Viner, Leslie, Kahn, Mason,
  Davis, Crannell, Hobbs, Schneeberger, Worden, Schommer, Vogt, Pettersen,
  Coleman, Karpen, Giampapa, Hege, Pazzani, Rodono, Romeo, \&
  Chugainov}]{kahler_coordinated_1982}
Kahler, S., Golub, L., Harnden, F.~R., {et~al.} 1982, The Astrophysical
  Journal, 252, 239.
\newblock \url{https://ui.adsabs.harvard.edu/abs/1982ApJ...252..239K}

\bibitem[{Kanodia {et~al.}(2019)Kanodia, Wolfgang, Stefansson, Ning, \&
  Mahadevan}]{kanodia_mass-radius_2019}
Kanodia, S., Wolfgang, A., Stefansson, G.~K., Ning, B., \& Mahadevan, S. 2019,
  The Astrophysical Journal, 882, 38.
\newblock \url{https://ui.adsabs.harvard.edu/abs/2019ApJ...882...38K}

\bibitem[{Kanodia \& Wright(2018)}]{kanodia_python_2018}
Kanodia, S., \& Wright, J. 2018, Research Notes of the AAS, 2, 4.
\newblock \url{http://stacks.iop.org/2515-5172/2/i=1/a=4}

\bibitem[{Kanodia {et~al.}(2018)Kanodia, Mahadevan, Ramsey, Stefansson, Monson,
  Hearty, Blakeslee, Lubar, Bender, Ninan, Sterner, Roy, Halverson, \&
  Robertson}]{kanodia_overview_2018}
Kanodia, S., Mahadevan, S., Ramsey, L.~W., {et~al.} 2018, SPIE Proceedings,
  0702, 107026Q, conference Name: Ground-based and Airborne Instrumentation for
  Astronomy VII ISBN: 9781510619579 Place: eprint: arXiv:1808.00557.
\newblock \url{http://adsabs.harvard.edu/abs/2018SPIE10702E..6QK}

\bibitem[{Kanodia {et~al.}(2021)Kanodia, Halverson, Ninan, Mahadevan,
  Stefansson, Roy, Ramsey, Bender, Janowiecki, Cochran, Diddams, Drory, Endl,
  Ford, Hearty, Metcalf, Monson, Robertson, Schwab, Terrien, \&
  Wright}]{kanodia_harsh_2021}
Kanodia, S., Halverson, S., Ninan, J.~P., {et~al.} 2021, The Astrophysical
  Journal, 912, 15.
\newblock \url{https://ui.adsabs.harvard.edu/abs/2021ApJ...912...15K}

\bibitem[{Kirkpatrick {et~al.}(1995)Kirkpatrick, Henry, \&
  Simons}]{kirkpatrick_solar_1995}
Kirkpatrick, J.~D., Henry, T.~J., \& Simons, D.~A. 1995, The Astronomical
  Journal, 109, 797.
\newblock \url{https://ui.adsabs.harvard.edu/abs/1995AJ....109..797K}

\bibitem[{Kotani {et~al.}(2018)Kotani, Tamura, Nishikawa, Ueda, Kuzuhara,
  Omiya, Hashimoto, Ishizuka, Hirano, Suto, Kurokawa, Kokubo, Mori, Tanaka,
  Kashiwagi, Konishi, Kudo, Sato, Jacobson, Hodapp, Hall, Aoki, Usuda,
  Nishiyama, Nakajima, Ikeda, Yamamuro, Morino, Baba, Hosokawa, Ishikawa,
  Narita, Kokubo, Hayano, Izumiura, Kambe, Kusakabe, Kwon, Ikoma, Hori, Genda,
  Fukui, Fujii, Kawahara, Olivier, Jovanovic, Harakawa, Hayashi, Hidai,
  Machida, Matsuo, Nagata, Ogihara, Takami, Takato, Terada, \&
  Oh}]{kotani_infrared_2018}
Kotani, T., Tamura, M., Nishikawa, J., {et~al.} 2018, SPIE, 0702, 1070211,
  conference Name: Ground-based and Airborne Instrumentation for Astronomy VII
  ISBN: 9781510619579.
\newblock \url{http://adsabs.harvard.edu/abs/2018SPIE10702E..11K}

\bibitem[{Kowalski {et~al.}(2015)Kowalski, Hawley, Carlsson, Allred,
  Uitenbroek, Osten, \& Holman}]{kowalski_new_2015}
Kowalski, A.~F., Hawley, S.~L., Carlsson, M., {et~al.} 2015, Solar Physics,
  290, 3487.
\newblock \url{https://ui.adsabs.harvard.edu/abs/2015SoPh..290.3487K}

\bibitem[{Kowalski {et~al.}(2013)Kowalski, Hawley, Wisniewski, Osten, Hilton,
  Holtzman, Schmidt, \& Davenport}]{kowalski_time-resolved_2013}
Kowalski, A.~F., Hawley, S.~L., Wisniewski, J.~P., {et~al.} 2013, The
  Astrophysical Journal Supplement Series, 207, 15.
\newblock \url{http://adsabs.harvard.edu/abs/2013ApJS..207...15K}

\bibitem[{Kowalski {et~al.}(2019)Kowalski, Wisniewski, Hawley, Osten, Brown,
  Fariña, Valenti, Brown, Xilouris, Schmidt, \&
  Johns-Krull}]{kowalski_near-ultraviolet_2019}
Kowalski, A.~F., Wisniewski, J.~P., Hawley, S.~L., {et~al.} 2019, The
  Astrophysical Journal, 871, 167, aDS Bibcode: 2019ApJ...871..167K.
\newblock \url{https://ui.adsabs.harvard.edu/abs/2019ApJ...871..167K}

\bibitem[{Kramida(2010)}]{kramida_critical_2010}
Kramida, A.~E. 2010, Atomic Data and Nuclear Data Tables, 96, 586.
\newblock
  \url{https://www.sciencedirect.com/science/article/pii/S0092640X10000458}

\bibitem[{Kramida {et~al.}(2020)Kramida, {Yu. Ralchenko}, {J. Reader}, \& {and
  NIST ASD Team}}]{kramida_notitle_2020}
Kramida, A.~E., {Yu. Ralchenko}, {J. Reader}, \& {and NIST ASD Team}. 2020,
  published: NIST Atomic Spectra Database (ver. 5.8), [Online]. Available:
  {\textbackslash}tthttps://physics.nist.gov/asd [2021, June 15]. National
  Institute of Standards and Technology, Gaithersburg, MD.

\bibitem[{Kunkel(1970)}]{kunkel_spectra_1970}
Kunkel, W.~E. 1970, The Astrophysical Journal, 161, 503.
\newblock \url{https://ui.adsabs.harvard.edu/abs/1970ApJ...161..503K}

\bibitem[{Lacatus {et~al.}(2017)Lacatus, Judge, \&
  Donea}]{lacatus_explanation_2017}
Lacatus, D.~A., Judge, P.~G., \& Donea, A. 2017, The Astrophysical Journal,
  842, 15.
\newblock \url{https://ui.adsabs.harvard.edu/abs/2017ApJ...842...15L}

\bibitem[{Lacy {et~al.}(1976)Lacy, Moffett, \& Evans}]{lacy_uv_1976}
Lacy, C.~H., Moffett, T.~J., \& Evans, D.~S. 1976, The Astrophysical Journal
  Supplement Series, 30, 85, aDS Bibcode: 1976ApJS...30...85L.
\newblock \url{https://ui.adsabs.harvard.edu/abs/1976ApJS...30...85L}

\bibitem[{Landman \& Illing(1977)}]{landman_measurements_1977}
Landman, D.~A., \& Illing, R. M.~E. 1977, Astronomy and Astrophysics, Vol. 55,
  p. 103 (1977), 55, 103.
\newblock
  \url{https://ui.adsabs.harvard.edu/abs/1977A%26A....55..103L/abstract}

\bibitem[{Lang {et~al.}(2010)Lang, Hogg, Mierle, Blanton, \&
  Roweis}]{lang_astrometrynet_2010}
Lang, D., Hogg, D.~W., Mierle, K., Blanton, M., \& Roweis, S. 2010, The
  Astronomical Journal, 139, 1782.
\newblock \url{http://adsabs.harvard.edu/abs/2010AJ....139.1782L}

\bibitem[{Li {et~al.}(2018)Li, Zhong, Xu, He, Yan, Chen, \&
  Fang}]{li_waiting_2018}
Li, C., Zhong, S.~J., Xu, Z.~G., {et~al.} 2018, Monthly Notices of the Royal
  Astronomical Society, 479, L139.
\newblock \url{https://ui.adsabs.harvard.edu/abs/2018MNRAS.479L.139L}

\bibitem[{Liebert {et~al.}(1999)Liebert, Kirkpatrick, Reid, \&
  Fisher}]{liebert_2mass_1999}
Liebert, J., Kirkpatrick, J.~D., Reid, I.~N., \& Fisher, M.~D. 1999, The
  Astrophysical Journal, 519, 345.
\newblock \url{http://adsabs.harvard.edu/abs/1999ApJ...519..345L}

\bibitem[{Liebert {et~al.}(1978)Liebert, Kron, \&
  Spinrad}]{liebert_spectrophotometry_1978}
Liebert, J., Kron, R.~G., \& Spinrad, H. 1978, Publications of the Astronomical
  Society of the Pacific, 90, 718.
\newblock \url{http://adsabs.harvard.edu/abs/1978PASP...90..718L}

\bibitem[{Linsky {et~al.}(1995)Linsky, Wood, Brown, Giampapa, \&
  Ambruster}]{linsky_stellar_1995}
Linsky, J.~L., Wood, B.~E., Brown, A., Giampapa, M.~S., \& Ambruster, C. 1995,
  The Astrophysical Journal, 455, 670.
\newblock \url{http://adsabs.harvard.edu/abs/1995ApJ...455..670L}

\bibitem[{Lockwood {et~al.}(2014)Lockwood, Johnson, Bender, Carr, Barman,
  Richert, \& Blake}]{lockwood_near-ir_2014}
Lockwood, A.~C., Johnson, J.~A., Bender, C.~F., {et~al.} 2014, The
  Astrophysical Journal, 783, L29.
\newblock \url{https://ui.adsabs.harvard.edu/abs/2014ApJ...783L..29L}

\bibitem[{Lopez \& Fortney(2014)}]{lopez_understanding_2014}
Lopez, E.~D., \& Fortney, J.~J. 2014, The Astrophysical Journal, 792, 1.
\newblock \url{https://ui.adsabs.harvard.edu/#abs/2014ApJ...792....1L/abstract}

\bibitem[{Luger {et~al.}(2015)Luger, Barnes, Lopez, Fortney, Jackson, \&
  Meadows}]{luger_habitable_2015}
Luger, R., Barnes, R., Lopez, E., {et~al.} 2015, Astrobiology, 15, 57.
\newblock \url{https://ui.adsabs.harvard.edu/abs/2015AsBio..15...57L}

\bibitem[{Mahadevan {et~al.}(2012)Mahadevan, Ramsey, Bender, Terrien, Wright,
  Halverson, Hearty, Nelson, Burton, Redman, Osterman, Diddams, Kasting, Endl,
  \& Deshpande}]{mahadevan_habitable-zone_2012}
Mahadevan, S., Ramsey, L., Bender, C., {et~al.} 2012, SPIE, 8446, 84461S,
  conference Name: Ground-based and Airborne Instrumentation for Astronomy IV
  Place: eprint: arXiv:1209.1686.
\newblock \url{https://ui.adsabs.harvard.edu/abs/2012SPIE.8446E..1SM}

\bibitem[{Mahadevan {et~al.}(2014)Mahadevan, Ramsey, Terrien, Halverson, Roy,
  Hearty, Levi, Stefansson, Robertson, Bender, Schwab, \&
  Nelson}]{mahadevan_habitable-zone_2014}
Mahadevan, S., Ramsey, L.~W., Terrien, R., {et~al.} 2014, SPIE, 9147, 91471G.
\newblock \url{http://adsabs.harvard.edu/abs/2014SPIE.9147E..1GM}

\bibitem[{Markwardt(2009)}]{markwardt_non-linear_2009}
Markwardt, C.~B. 2009, 411, 251, conference Name: Astronomical Data Analysis
  Software and Systems XVIII Place: eprint: arXiv:0902.2850.
\newblock \url{https://ui.adsabs.harvard.edu/abs/2009ASPC..411..251M}

\bibitem[{Martin {et~al.}(2017)Martin, Fuhrmeister, Mittag, Schmidt,
  Hempelmann, González-Pérez, \& Schmitt}]{martin_ca_2017}
Martin, J., Fuhrmeister, B., Mittag, M., {et~al.} 2017, Astronomy and
  Astrophysics, 605, A113.
\newblock \url{http://adsabs.harvard.edu/abs/2017A%26A...605A.113M}

\bibitem[{Martín(1999)}]{martin_utrecht_1999}
Martín, E.~L. 1999, Monthly Notices of the Royal Astronomical Society, 302,
  59.
\newblock \url{https://ui.adsabs.harvard.edu/abs/1999MNRAS.302...59M}

\bibitem[{McDonald {et~al.}(2019)McDonald, Kreidberg, \&
  Lopez}]{mcdonald_sub-neptune_2019}
McDonald, G.~D., Kreidberg, L., \& Lopez, E. 2019, The Astrophysical Journal,
  876, 22.
\newblock \url{http://adsabs.harvard.edu/abs/2019ApJ...876...22M}

\bibitem[{McKinney(2010)}]{mckinney_data_2010}
McKinney, W. 2010, in Proceedings of the 9th {Python} in {Science}
  {Conference}, ed. S.~v.~d. Walt \& J.~Millman, 56 -- 61

\bibitem[{Metcalf {et~al.}(2019)Metcalf, Anderson, Bender, Blakeslee, Brand,
  Carlson, Cochran, Diddams, Endl, Fredrick, Halverson, Hickstein, Hearty,
  Jennings, Kanodia, Kaplan, Levi, Lubar, Mahadevan, Monson, Ninan, Nitroy,
  Osterman, Papp, Quinlan, Ramsey, Robertson, Roy, Schwab, Sigurdsson,
  Srinivasan, Stefansson, Sterner, Terrien, Wolszczan, Wright, \&
  Ycas}]{metcalf_stellar_2019}
Metcalf, A.~J., Anderson, T., Bender, C.~F., {et~al.} 2019, Optica, 6, 233.
\newblock \url{https://ui.adsabs.harvard.edu/abs/2019Optic...6..233M}

\bibitem[{Mochnacki \& Zirin(1980)}]{mochnacki_multichannel_1980}
Mochnacki, S.~W., \& Zirin, H. 1980, The Astrophysical Journal, 239, L27.
\newblock \url{https://ui.adsabs.harvard.edu/abs/1980ApJ...239L..27M}

\bibitem[{Mohanty \& Basri(2003)}]{mohanty_rotation_2003}
Mohanty, S., \& Basri, G. 2003, The Astrophysical Journal, 583, 451.
\newblock \url{https://ui.adsabs.harvard.edu/abs/2003ApJ...583..451M}

\bibitem[{Montes {et~al.}(2000)Montes, Fernández-Figueroa, De~Castro, Cornide,
  Latorre, \& Sanz-Forcada}]{montes_multiwavelength_2000}
Montes, D., Fernández-Figueroa, M.~J., De~Castro, E., {et~al.} 2000, Astronomy
  and Astrophysics Supplement, v.146, p.103-140, 146, 103.
\newblock
  \url{https://ui.adsabs.harvard.edu/abs/2000A%26AS..146..103M/abstract}

\bibitem[{Muheki {et~al.}(2020)Muheki, Guenther, Mutabazi, \&
  Jurua}]{muheki_high-resolution_2020}
Muheki, P., Guenther, E.~W., Mutabazi, T., \& Jurua, E. 2020, Astronomy and
  Astrophysics, 637, A13.
\newblock \url{https://ui.adsabs.harvard.edu/abs/2020A&A...637A..13M/abstract}

\bibitem[{Mulders {et~al.}(2015)Mulders, Pascucci, \&
  Apai}]{mulders_stellar-mass-dependent_2015}
Mulders, G.~D., Pascucci, I., \& Apai, D. 2015, The Astrophysical Journal, 798,
  112.
\newblock \url{http://adsabs.harvard.edu/abs/2015ApJ...798..112M}

\bibitem[{National(1992)}]{national_us_1992}
National, C. G.~D. 1992, {\textbackslash}planss, 40, 553

\bibitem[{Neidig \& Wiborg(1984)}]{neidig_hydrogen_1984}
Neidig, D.~F., \& Wiborg, Jr., P.~H. 1984, Solar Physics, 92, 217.
\newblock \url{http://adsabs.harvard.edu/abs/1984SoPh...92..217N}

\bibitem[{Ninan {et~al.}(2018)Ninan, Bender, Mahadevan, Ford, Monson, Kaplan,
  Terrien, Roy, Robertson, Kanodia, \& Stefansson}]{ninan_habitable-zone_2018}
Ninan, J.~P., Bender, C.~F., Mahadevan, S., {et~al.} 2018, 0709, 107092U.
\newblock \url{http://adsabs.harvard.edu/abs/2018SPIE10709E..2UN}

\bibitem[{Ninan {et~al.}(2020)Ninan, Stefansson, Mahadevan, Bender, Robertson,
  Ramsey, Terrien, Wright, Diddams, Kanodia, Cochran, Endl, Ford, Fredrick,
  Halverson, Hearty, Jennings, Kaplan, Lubar, Metcalf, Monson, Nitroy, Roy, \&
  Schwab}]{ninan_evidence_2020}
Ninan, J.~P., Stefansson, G., Mahadevan, S., {et~al.} 2020, The Astrophysical
  Journal, 894, 97.
\newblock \url{http://adsabs.harvard.edu/abs/2020ApJ...894...97N}

\bibitem[{Oliphant(2006)}]{oliphant_numpy_2006}
Oliphant, T. 2006, {NumPy}: {A} guide to {NumPy}, published: USA: Trelgol
  Publishing.
\newblock \url{http://www.numpy.org/}

\bibitem[{Oliphant(2007)}]{oliphant_python_2007}
Oliphant, T.~E. 2007, Computing in Science Engineering, 9, 10

\bibitem[{Osten {et~al.}(2012)Osten, Kowalski, Sahu, \&
  Hawley}]{osten_drafts_2012}
Osten, R.~A., Kowalski, A., Sahu, K., \& Hawley, S.~L. 2012, The Astrophysical
  Journal, 754, 4, aDS Bibcode: 2012ApJ...754....4O.
\newblock \url{https://ui.adsabs.harvard.edu/abs/2012ApJ...754....4O}

\bibitem[{Osterbrock(1989)}]{osterbrock_astrophysics_1989}
Osterbrock, D.~E. 1989, Astrophysics of Gaseous Nebulae and Active Galactic
  Nuclei.
\newblock \url{https://ui.adsabs.harvard.edu/abs/1989agna.book.....O}

\bibitem[{Owen \& Jackson(2012)}]{owen_planetary_2012}
Owen, J.~E., \& Jackson, A.~P. 2012, Monthly Notices of the Royal Astronomical
  Society, 425, 2931.
\newblock \url{https://ui.adsabs.harvard.edu/abs/2012MNRAS.425.2931O}

\bibitem[{Owen \& Mohanty(2016)}]{owen_habitability_2016}
Owen, J.~E., \& Mohanty, S. 2016, Monthly Notices of the Royal Astronomical
  Society, 459, 4088.
\newblock \url{https://ui.adsabs.harvard.edu/abs/2016MNRAS.459.4088O}

\bibitem[{Paschen(1908)}]{paschen_zur_1908}
Paschen, F. 1908, Annalen der Physik, 332, 537, \_eprint:
  https://onlinelibrary.wiley.com/doi/pdf/10.1002/andp.19083321303.
\newblock
  \url{https://onlinelibrary.wiley.com/doi/abs/10.1002/andp.19083321303}

\bibitem[{Paudel {et~al.}(2018)Paudel, Gizis, Mullan, Schmidt, Burgasser,
  Williams, \& Berger}]{paudel_k2_2018}
Paudel, R.~R., Gizis, J.~E., Mullan, D.~J., {et~al.} 2018, The Astrophysical
  Journal, 858, 55.
\newblock \url{https://ui.adsabs.harvard.edu/abs/2018ApJ...858...55P}

\bibitem[{Paudel {et~al.}(2019)Paudel, Gizis, Mullan, Schmidt, Burgasser,
  Williams, Youngblood, \& Stassun}]{paudel_k2_2019}
---. 2019, Monthly Notices of the Royal Astronomical Society, 486, 1438.
\newblock \url{https://ui.adsabs.harvard.edu/abs/2019MNRAS.486.1438P}

\bibitem[{Paulson {et~al.}(2006)Paulson, Allred, Anderson, Hawley, Cochran, \&
  Yelda}]{paulson_optical_2006}
Paulson, D.~B., Allred, J.~C., Anderson, R.~B., {et~al.} 2006, Publications of
  the Astronomical Society of the Pacific, 118, 227.
\newblock \url{http://adsabs.harvard.edu/abs/2006PASP..118..227P}

\bibitem[{Penz \& Micela(2008)}]{penz_x-ray_2008}
Penz, T., \& Micela, G. 2008, Astronomy and Astrophysics, 479, 579.
\newblock \url{https://ui.adsabs.harvard.edu/abs/2008A&A...479..579P/abstract}

\bibitem[{Pettersen \& Hawley(1989)}]{pettersen_spectroscopic_1989}
Pettersen, B.~R., \& Hawley, S.~L. 1989, Astronomy and Astrophysics, Vol. 217,
  p. 187-200 (1989), 217, 187.
\newblock
  \url{https://ui.adsabs.harvard.edu/abs/1989A%26A...217..187P/abstract}

\bibitem[{Pérez \& Granger(2007)}]{perez_ipython_2007}
Pérez, F., \& Granger, B.~E. 2007, Computing in Science and Engineering, 9,
  21.
\newblock \url{https://ipython.org}

\bibitem[{Quirrenbach {et~al.}(2014)Quirrenbach, Amado, Caballero, Mundt,
  Reiners, Ribas, Seifert, Abril, Aceituno, Alonso-Floriano, Ammler-von Eiff,
  Antona~Jiménez, Anwand-Heerwart, Azzaro, Bauer, Barrado, Becerril, Béjar,
  Benítez, Berdiñas, Cárdenas, Casal, Claret, Colomé, Cortés-Contreras,
  Czesla, Doellinger, Dreizler, Feiz, Fernández, Galadí, Gálvez-Ortiz,
  García-Piquer, García-Vargas, Garrido, Gesa, Gómez~Galera,
  González~Álvarez, González~Hernández, Grözinger, Guàrdia, Guenther,
  de~Guindos, Gutiérrez-Soto, Hagen, Hatzes, Hauschildt, Helmling, Henning,
  Hermann, Hernández~Castaño, Herrero, Hidalgo, Holgado, Huber, Huber,
  Jeffers, Joergens, de~Juan, Kehr, Klein, Kürster, Lamert, Lalitha, Laun,
  Lemke, Lenzen, López~del Fresno, López~Martí, López-Santiago, Mall,
  Mandel, Martín, Martín-Ruiz, Martínez-Rodríguez, Marvin, Mathar, Mirabet,
  Montes, Morales~Muñoz, Moya, Naranjo, Ofir, Oreiro, Pallé, Panduro,
  Passegger, Pérez-Calpena, Pérez~Medialdea, Perger, Pluto, Ramón, Rebolo,
  Redondo, Reffert, Reinhardt, Rhode, Rix, Rodler, Rodríguez,
  Rodríguez-López, Rodríguez-Pérez, Rohloff, Rosich, Sánchez-Blanco,
  Sánchez~Carrasco, Sanz-Forcada, Sarmiento, Schäfer, Schiller, Schmidt,
  Schmitt, Solano, Stahl, Storz, Stürmer, Suárez, Ulbrich, Veredas, Wagner,
  Winkler, Zapatero~Osorio, Zechmeister, Abellán~de Paco, Anglada-Escudé, del
  Burgo, Klutsch, Lizon, López-Morales, Morales, Perryman, Tulloch, \&
  Xu}]{quirrenbach_carmenes_2014}
Quirrenbach, A., Amado, P.~J., Caballero, J.~A., {et~al.} 2014, 9147, 91471F,
  conference Name: Ground-based and Airborne Instrumentation for Astronomy V.
\newblock \url{https://ui.adsabs.harvard.edu/abs/2014SPIE.9147E..1FQ}

\bibitem[{Ramsay {et~al.}(2021)Ramsay, Kolotkov, Doyle, \&
  Doyle}]{ramsay_tess_2021}
Ramsay, G., Kolotkov, D., Doyle, J.~G., \& Doyle, L. 2021, arXiv:2108.10670
  [astro-ph], arXiv: 2108.10670.
\newblock \url{http://arxiv.org/abs/2108.10670}

\bibitem[{Ramsey {et~al.}(1998)Ramsey, Adams, Barnes, Booth, Cornell, Fowler,
  Gaffney, Glaspey, Good, Hill, Kelton, Krabbendam, Long, MacQueen, Ray,
  Ricklefs, Sage, Sebring, Spiesman, \& Steiner}]{ramsey_early_1998}
Ramsey, L.~W., Adams, M.~T., Barnes, T.~G., {et~al.} 1998, 3352, 34, conference
  Name: Advanced Technology Optical/IR Telescopes VI.
\newblock \url{http://adsabs.harvard.edu/abs/1998SPIE.3352...34R}

\bibitem[{Reid \& Hawley(2000)}]{reid_new_2000}
Reid, I.~N., \& Hawley, S.~L. 2000, New light on dark stars. Red dwarfs.
\newblock \url{https://ui.adsabs.harvard.edu/abs/2000nlod.book.....R}

\bibitem[{Reiners {et~al.}(2018)Reiners, Zechmeister, Caballero, Ribas,
  Morales, Jeffers, Schöfer, Tal-Or, Quirrenbach, Amado, Kaminski, Seifert,
  Abril, Aceituno, Alonso-Floriano, Ammler-von Eiff, Antona, Anglada-Escudé,
  Anwand-Heerwart, Arroyo-Torres, Azzaro, Baroch, Barrado, Bauer, Becerril,
  Béjar, Benítez, Berdinas̃, Bergond, Blümcke, Brinkmöller, del Burgo,
  Cano, Cárdenas~Vázquez, Casal, Cifuentes, Claret, Colomé,
  Cortés-Contreras, Czesla, Díez-Alonso, Dreizler, Feiz, Fernández, Ferro,
  Fuhrmeister, Galadí-Enríquez, Garcia-Piquer, García~Vargas, Gesa,
  Gómez~Galera, González~Hernández, González-Peinado, Grözinger, Grohnert,
  Guàrdia, Guenther, Guijarro, de~Guindos, Gutiérrez-Soto, Hagen, Hatzes,
  Hauschildt, Hedrosa, Helmling, Henning, Hermelo, Hernández~Arabí,
  Hernández~Castaño, Hernández~Hernando, Herrero, Huber, Huke, Johnson,
  de~Juan, Kim, Klein, Klüter, Klutsch, Kürster, Lafarga, Lamert, Lampón,
  Lara, Laun, Lemke, Lenzen, Launhardt, López~del Fresno, López-González,
  López-Puertas, López~Salas, López-Santiago, Luque, Magán~Madinabeitia,
  Mall, Mancini, Mandel, Marfil, Marín~Molina, Maroto~Fernández, Martín,
  Martín-Ruiz, Marvin, Mathar, Mirabet, Montes, Moreno-Raya, Moya, Mundt,
  Nagel, Naranjo, Nortmann, Nowak, Ofir, Oreiro, Pallé, Panduro, Pascual,
  Passegger, Pavlov, Pedraz, Pérez-Calpena, Pérez~Medialdea, Perger,
  Perryman, Pluto, Rabaza, Ramón, Rebolo, Redondo, Reffert, Reinhart, Rhode,
  Rix, Rodler, Rodríguez, Rodríguez-López, Rodríguez~Trinidad, Rohloff,
  Rosich, Sadegi, Sánchez-Blanco, Sánchez~Carrasco, Sánchez-López,
  Sanz-Forcada, Sarkis, Sarmiento, Schäfer, Schmitt, Schiller, Schweitzer,
  Solano, Stahl, Strachan, Stürmer, Suárez, Tabernero, Tala, Trifonov,
  Tulloch, Ulbrich, Veredas, Vico~Linares, Vilardell, Wagner, Winkler,
  Wolthoff, Xu, Yan, \& Zapatero~Osorio}]{reiners_carmenes_2018}
Reiners, A., Zechmeister, M., Caballero, J.~A., {et~al.} 2018, Astronomy \&
  Astrophysics, 612, A49.
\newblock \url{https://www.aanda.org/10.1051/0004-6361/201732054}

\bibitem[{Ricker {et~al.}(2014)Ricker, Winn, Vanderspek, Latham, Bakos, Bean,
  Berta-Thompson, Brown, Buchhave, Butler, Butler, Chaplin, Charbonneau,
  Christensen-Dalsgaard, Clampin, Deming, Doty, Lee, Dressing, Dunham, Endl,
  Fressin, Ge, Henning, Holman, Howard, Ida, Jenkins, Jernigan, Johnson,
  Kaltenegger, Kawai, Kjeldsen, Laughlin, Levine, Lin, Lissauer, MacQueen,
  Marcy, McCullough, Morton, Narita, Paegert, Palle, Pepe, Pepper, Quirrenbach,
  Rinehart, Sasselov, Sato, Seager, Sozzetti, Stassun, Sullivan, Szentgyorgyi,
  Torres, Udry, \& Villasenor}]{ricker_transiting_2014}
Ricker, G.~R., Winn, J.~N., Vanderspek, R., {et~al.} 2014, Journal of
  Astronomical Telescopes, Instruments, and Systems, 1, 014003.
\newblock
  \url{https://www.spiedigitallibrary.org/journals/Journal-of-Astronomical-Telescopes-Instruments-and-Systems/volume-1/issue-1/014003/Transiting-Exoplanet-Survey-Satellite/10.1117/1.JATIS.1.1.014003.short}

\bibitem[{Robertson {et~al.}(2016)Robertson, Bender, Mahadevan, Roy, \&
  Ramsey}]{robertson_proxima_2016}
Robertson, P., Bender, C., Mahadevan, S., Roy, A., \& Ramsey, L.~W. 2016, The
  Astrophysical Journal, 832, 112.
\newblock \url{http://adsabs.harvard.edu/abs/2016ApJ...832..112R}

\bibitem[{Robitaille {et~al.}(2013)Robitaille, Tollerud, Greenfield,
  Droettboom, Bray, Aldcroft, Davis, Ginsburg, Price-Whelan, Kerzendorf,
  Conley, Crighton, Barbary, Muna, Ferguson, Grollier, Parikh, Nair, Günther,
  Deil, Woillez, Conseil, Kramer, Turner, Singer, Fox, Weaver, Zabalza,
  Edwards, Bostroem, Burke, Casey, Crawford, Dencheva, Ely, Jenness, Labrie,
  Lim, Pierfederici, Pontzen, Ptak, Refsdal, Servillat, \&
  Streicher}]{robitaille_astropy_2013}
Robitaille, T.~P., Tollerud, E.~J., Greenfield, P., {et~al.} 2013, Astronomy \&
  Astrophysics, 558, A33.
\newblock
  \url{https://www.aanda.org/articles/aa/abs/2013/10/aa22068-13/aa22068-13.html}

\bibitem[{Rothman {et~al.}(2013)Rothman, Gordon, Babikov, Barbe, Chris~Benner,
  Bernath, Birk, Bizzocchi, Boudon, Brown, Campargue, Chance, Cohen, Coudert,
  Devi, Drouin, Fayt, Flaud, Gamache, Harrison, Hartmann, Hill, Hodges,
  Jacquemart, Jolly, Lamouroux, Le~Roy, Li, Long, Lyulin, Mackie, Massie,
  Mikhailenko, Müller, Naumenko, Nikitin, Orphal, Perevalov, Perrin,
  Polovtseva, Richard, Smith, Starikova, Sung, Tashkun, Tennyson, Toon,
  Tyuterev, \& Wagner}]{rothman_hitran2012_2013}
Rothman, L.~S., Gordon, I.~E., Babikov, Y., {et~al.} 2013, Journal of
  Quantitative Spectroscopy and Radiative Transfer, 130, 4.
\newblock \url{https://ui.adsabs.harvard.edu/abs/2013JQSRT.130....4R/abstract}

\bibitem[{Ruan {et~al.}(2021)Ruan, Zhou, \& Keppens}]{ruan_when_2021}
Ruan, W., Zhou, Y., \& Keppens, R. 2021, The Astrophysical Journal Letters,
  920, L15, arXiv: 2109.11873.
\newblock \url{http://arxiv.org/abs/2109.11873}

\bibitem[{Scalo {et~al.}(2007)Scalo, Kaltenegger, Segura, Fridlund, Ribas,
  Kulikov, Grenfell, Rauer, Odert, Leitzinger, Selsis, Khodachenko, Eiroa,
  Kasting, \& Lammer}]{scalo_m_2007}
Scalo, J., Kaltenegger, L., Segura, A., {et~al.} 2007, Astrobiology, 7, 85.
\newblock \url{https://www.liebertpub.com/doi/10.1089/ast.2006.0125}

\bibitem[{Schmidt {et~al.}(2007)Schmidt, Cruz, Bongiorno, Liebert, \&
  Reid}]{schmidt_activity_2007}
Schmidt, S.~J., Cruz, K.~L., Bongiorno, B.~J., Liebert, J., \& Reid, I.~N.
  2007, The Astronomical Journal, 133, 2258.
\newblock \url{http://adsabs.harvard.edu/abs/2007AJ....133.2258S}

\bibitem[{Schmidt {et~al.}(2012)Schmidt, Kowalski, Hawley, Hilton, Wisniewski,
  \& Tofflemire}]{schmidt_probing_2012}
Schmidt, S.~J., Kowalski, A.~F., Hawley, S.~L., {et~al.} 2012, The
  Astrophysical Journal, 745, 14.
\newblock \url{http://adsabs.harvard.edu/abs/2012ApJ...745...14S}

\bibitem[{Schöfer {et~al.}(2019)Schöfer, Jeffers, Reiners, Shulyak,
  Fuhrmeister, Johnson, Zechmeister, Ribas, Quirrenbach, Amado, Caballero,
  Anglada-Escudé, Bauer, Béjar, Cortés-Contreras, Dreizler, Guenther,
  Kaminski, Kürster, Lafarga, Montes, Morales, Pedraz, \&
  Tal-Or}]{schofer_carmenes_2019}
Schöfer, P., Jeffers, S.~V., Reiners, A., {et~al.} 2019, Astronomy \&
  Astrophysics, 623, A44.
\newblock \url{https://www.aanda.org/10.1051/0004-6361/201834114}

\bibitem[{Seifahrt {et~al.}(2016)Seifahrt, Bean, Stürmer, Gers, Grobler, Reed,
  \& Jones}]{seifahrt_development_2016}
Seifahrt, A., Bean, J.~L., Stürmer, J., {et~al.} 2016, in Ground-based and
  {Airborne} {Instrumentation} for {Astronomy} {VI}, Vol. 9908 (International
  Society for Optics and Photonics), 990818.
\newblock
  \url{https://www.spiedigitallibrary.org/conference-proceedings-of-spie/9908/990818/Development-and-construction-of-MAROON-X/10.1117/12.2232069.short}

\bibitem[{Stefansson {et~al.}(2016)Stefansson, Hearty, Robertson, Mahadevan,
  Anderson, Levi, Bender, Nelson, Monson, Blank, Halverson, Henderson, Ramsey,
  Roy, Schwab, \& Terrien}]{stefansson_versatile_2016}
Stefansson, G., Hearty, F., Robertson, P., {et~al.} 2016, The Astrophysical
  Journal, 833, 175.
\newblock \url{http://adsabs.harvard.edu/abs/2016ApJ...833..175S}

\bibitem[{Tinney {et~al.}(1993)Tinney, Reid, \& Mould}]{tinney_faintest_1993}
Tinney, C.~G., Reid, I.~N., \& Mould, J.~R. 1993, The Astrophysical Journal,
  414, 254.
\newblock \url{https://ui.adsabs.harvard.edu/abs/1993ApJ...414..254T}

\bibitem[{Valenti {et~al.}(1995)Valenti, Butler, \&
  Marcy}]{valenti_determining_1995}
Valenti, J.~A., Butler, R.~P., \& Marcy, G.~W. 1995, Publications of the
  Astronomical Society of the Pacific, 107, 966.
\newblock \url{https://ui.adsabs.harvard.edu/abs/1995PASP..107..966V}

\bibitem[{van Biesbroeck(1944)}]{van_biesbroeck_star_1944}
van Biesbroeck, G. 1944, The Astronomical Journal, 51, 61.
\newblock \url{http://adsabs.harvard.edu/abs/1944AJ.....51...61V}

\bibitem[{van Dokkum(2001)}]{van_dokkum_cosmic-ray_2001}
van Dokkum, P.~G. 2001, Publications of the Astronomical Society of the
  Pacific, 113, 1420.
\newblock \url{http://adsabs.harvard.edu/abs/2001PASP..113.1420V}

\bibitem[{van Maanen(1940)}]{van_maanen_photographic_1940}
van Maanen, A. 1940, The Astrophysical Journal, 91, 503.
\newblock \url{https://ui.adsabs.harvard.edu/abs/1940ApJ....91..503V}

\bibitem[{Virtanen {et~al.}(2020)Virtanen, Gommers, Oliphant, Haberland, Reddy,
  Cournapeau, Burovski, Peterson, Weckesser, Bright, van~der Walt, Brett,
  Wilson, Jarrod~Millman, Mayorov, Nelson, Jones, Kern, Larson, Carey, Polat,
  Feng, Moore, Vand~erPlas, Laxalde, Perktold, Cimrman, Henriksen, Quintero,
  Harris, Archibald, Ribeiro, Pedregosa, van Mulbregt, \&
  Contributors}]{virtanen_scipy_2020}
Virtanen, P., Gommers, R., Oliphant, T.~E., {et~al.} 2020, Nature Methods, 17,
  261

\bibitem[{Wildi {et~al.}(2017)Wildi, Blind, Reshetov, Hernandez, Genolet,
  Conod, Sordet, Segovilla, Rasilla, Brousseau, Thibault, Delabre, Bandy,
  Sarajlic, Cabral, Bovay, Vallée, Bouchy, Doyon, Artigau, Pepe, Hagelberg,
  Melo, Delfosse, Figueira, Santos, González~Hernández, de~Medeiros, Rebolo,
  Broeg, Benz, Boisse, Malo, Käufl, \& Saddlemyer}]{wildi_nirps_2017}
Wildi, F., Blind, N., Reshetov, V., {et~al.} 2017, in , 1040018.
\newblock \url{http://adsabs.harvard.edu/abs/2017SPIE10400E..18W}

\bibitem[{Wright \& Eastman(2014)}]{wright_barycentric_2014}
Wright, J.~T., \& Eastman, J.~D. 2014, Publications of the Astronomical Society
  of the Pacific, 126, 838.
\newblock \url{https://ui.adsabs.harvard.edu/abs/2014PASP..126..838W}

\bibitem[{Zapatero~Osorio {et~al.}(2009)Zapatero~Osorio, Martín, del Burgo,
  Deshpande, Rodler, \& Montgomery}]{zapatero_osorio_infrared_2009}
Zapatero~Osorio, M.~R., Martín, E.~L., del Burgo, C., {et~al.} 2009, Astronomy
  and Astrophysics, 505, L5.
\newblock \url{http://adsabs.harvard.edu/abs/2009A%26A...505L...5Z}

\end{thebibliography}

\listofchanges
\end{document}